\documentclass[12pt]{article}
\usepackage[margin=1.2in]{geometry}

\usepackage{graphicx}
\usepackage[T1]{fontenc}
\usepackage{verbatimbox}

\usepackage{gensymb}
\usepackage{booktabs}
\usepackage{adjustbox}
\usepackage{lscape}
\usepackage{subcaption}
\usepackage{caption}
\usepackage{multirow}
\usepackage{hyperref}
\usepackage[colorinlistoftodos]{todonotes}

\usepackage{amssymb}
\usepackage[utf8]{inputenc}
\usepackage{color}
\usepackage{amsmath}

\usepackage{algorithm}
\usepackage{algpseudocode}

\usepackage{natbib}
\setcitestyle{authoryear,open={(},close={)}}

\newcommand{\ignore}[1]{}

\usepackage{setspace}
\begin{document}


\title{Cell2Fire: A Cell Based Forest Fire Growth Model}

\author{Cristobal Pais$^{a}$, Jaime Carrasco$^{b}$, \\
David L.\ Martell$^{c}$, Andres Weintraub$^{b}$, David L.\ Woodruff $^{d}$}

\maketitle         

\noindent
{$^{a}$University of California Berkeley, IEOR Department\\ 
         $^{b}$University of Chile, Industrial Engineering Department\\
         $^{c}$University of Toronto, Faculty of Forestry\\
         $^{d}$University of California Davis, Graduate School of Management}

\begin{abstract}
\noindent
Cell2Fire is a new cell-based forest and wildland landscape fire growth simulator that is open-source and exploits parallelism to support the modelling of fire growth cross large spatial and temporal scales in a timely manner.
The fire environment is characterized by partitioning the landscape into a large number of cells each of which has specified fuel, weather, fuel moisture and topography attributes.
Fire spread within each cell is assumed to be elliptical and governed by spread rates predicted by a fire spread model such as the Canadian Forest Fire Behavior Prediction (FBP) System.
The simulator includes powerful statistical and graphical output and spatial analysis features to facilitate the display and analysis of projected fire growth.

We validated Cell2Fire by using it to predict the growth of real and realistic hypothetical fires, comparing our fire growth predictions with those produced by the state-of-the-art Prometheus fire growth simulator. Cell2Fire is structured to facilitate its use for predicting the growth of individual fires or embedding it in landscape management simulation models. It can be used to produce probabilistic fire scar predictions by allowing for uncertainty concerning the basic spread rate predictions and uncertain weather scenarios that might drive their growth.

\end{abstract}

{\bf keywords:}
Forest fire spread, FireSmart forest management, Fire simulation, Wildfire, Cellular automata.


\section{Introduction}
\label{S:1}

The effects of global warming on temperature, precipitation levels, soil moisture and other forest and wildland fire regime drivers have increased and are expected to continue to increase both the number of and area burned by wildfires around the globe \citep{Westerling2016}. Wildfires have burned large areas and important infrastructure, thousands of homes and forest resources have been destroyed and many lives have been lost in recent years.
Recent examples include catastrophic incidents in the United States, Canada, Chile, Portugal and southwestern Australia in the years 2016-2018.
That has also resulted in increases in expenditures by forest and wildland fire management agencies \citep[see, e.g.,][]{CanadaMoney}.
Despite concerted efforts, wildfire growth remains a complex and very difficult to model process. 

Two of the most important characteristics of a wildfire are its rate of spread (ROS) and intensity which are influenced by fuel type, fuel moisture, wind velocity, and slope. The Canadian Forest Fire Behavior Prediction (FBP) System includes empirical fire spread rate models that can be used to predict the rate of spread and the intensity of wildfires based on weather, fuel moisture, time of year and topographical variables for specified fuel types; e.g., for individual grid cells that contain homogeneous fuel types \citep{FBP}. 

However, the FBP System alone cannot be used to predict how a fire will grow across a heterogeneous landscape/grid over time.
Spatial fire growth models like Prometheus, a deterministic fire growth simulator, are designed to use FBP spread rates to do so \citep{Prometheus}. 
Prometheus is a vector-based fire growth simulation model that is based on an adaptation of Huygens’ principle of wave propagation, i.e., the propagation of the fire front is modelled in a fashion similar to a wave, shifting and moving forward continuously in time and space.
It uses spatially explicit fire environment input data concerning topography (slope, aspect and elevation) and FBP fuel types along with a weather stream and fire danger rating codes and indices,  

FARSITE \cite{Finney2005} is another widely used fire growth simulator. It is based on the U.S.\ Forest Service's BEHAVE fire behavior prediction system and is also a vector-based Huygens' type model. A review of twenty-three simulators that can be used to predict forest fire growth can be found in \cite{SimReview}. The two models found to best simulate historical fires were FARSITE in the United States and Prometheus in Canada.

Recent years have witnessed growing interest in the development of detailed cell-based deterministic/stochastic fire simulators and some of the modelling assumptions that have been adopted include, for example the use of memoryless distributions (Markovian processes) to model the fire spread dynamics \citep{Boychuk2009}, homogeneous forests (cells' characteristics are identical), reductions in the number of adjacent cells to which a fire can spread (e.g., from 8 to 6 or 4), no spotting, and no stochasticity is included.

Our goal was to include many realistic aspects of fire behaviour in Cell2Fire --- an open-source (\url{https://github.com/cell2fire/Cell2Fire}) cell-based fire growth simulator, with a view to achieving high computational performance via parallelism when simulating large-scale fire instances to provide valuable insight to inform both fire and forest management. We therefore use both real and realistic hypothetical fire instances to validate our simulator and assess its computational performance.

FireSmart \citep{hirschetal} forest and fuel management calls for landscape management and fire growth simulation models that have well-structured interfaces that facilitate the exchange of data between them to inform the iterative re-planning that takes place when strategic plans are modified in response to what fires materialize and what areas actually burn over long planning horizons.

Acuna et al. (2010) demonstrated the importance of integrating fire management with strategic forest management planning models to develop and evaluate FireSmart forest management plans.  Although stand-alone fire growth models can be used to evaluate specified forest and fuel management plans, the fire growth simulators described above were designed to simulate fire growth and cannot readily be incorporated strategic planning frameworks that can be used to develop good or optimal strategic FireSmart landscape management plans.

Cell2Fire is designed specifically for use in a fuel-management framework with the intent of mitigating the detrimental impact of large fires efficiently. Cell2Fire can therefore be used as a pure simulation tool to model the growth of specific fires over a short time frame and/or embedded in a landscape management framework to evaluate fuel-management strategies or linked with optimization software to develop ``optimal'' fuel management strategies over long planning horizons.

The primary objective of our research was to develop an efficient and realistic fire growth simulator that enables simulation of the dynamics of fire growth across a grid representation of a real or hypothetical forest using cell attributes such as fuel type, elevation and weather, given an ignition point or initial fire perimeter, to inform FireSmart forest management. Our ultimate goal is to develop a tractable methodology that can be used to generate realistic spatial fire scar scenarios that can be used to support fuel management and harvesting planning.

The paper is organized as follows: In section 2, the two main fire growth simulation paradigms are described and a review of state-of-the-art simulators is included. Section 3 describes the Cell2Fire simulator structure, the main simulation steps, the fire growth dynamics model and the computational implementation. The results of a case study based on a real forest landscape in Canada and several test instances that were used to validate the simulator output and compare its computational performance with state-of-the-art simulators are discussed in section 4. Finally, section 5 contains our conclusions and thoughts concerning future research needs.

  
\section{Description of fire growth dynamics} 
The two  methods that have most often been used to simulate fire spread rates and fire growth across heterogeneous landscapes are the wave propagation approach and the cellular automata approach. We therefore begin by providing a brief overview of those two approaches.

\subsection{Wave-propagation model: Huygens}
Huygens considered every point on a wavefront of light as a source of individual wavelets and described the new wavefront as the surface tangential to the circumferences of the secondary waves. The use of Huygens’ principle to simulate fire growth is based on the assumption that the shape of a fire can be represented by a polygon, a plane figure composed of a sequence of straight-line segments forming a closed path, whose vertices are a tangential envelope of the elliptical ``firelets''. Huygens’ principle was first applied to the
simulation of fire spread by\ignore{Sanderlin and Sunderson in} \citet{Sanderlin1975}. \ignore{Anderson et al. in} \citet{Anderson19822} later developed a simple elliptical model based on Huygens’ principle of wave propagation to simulate the spread of grass fires. \ignore{ Richards in} \citet{Richards1990} then extended the  Anderson et al (1982) model by deriving a set of partial differential equations to model the growth of fires across a heterogeneous landscape.

Both the FARSITE \citep{Finney2004} and Prometheus models use\ignore{ Richards} \citet{Richards1990} partial differential equations to propagate each vertex on a fire's perimeter. However, the models differ in the fire danger rating system components and fuel models used to model fire spread rates. FARSITE uses the US National Fire Danger Rating System and fire behavior prediction fuel models developed by\ignore{Rothermel} \citet{Rothermel1972} and extended by\ignore{ Anderson} \citet{Anderson1982} and \ignore{Scott and Burgan }\citet{Scott&Burgan}, whereas Prometheus uses the Canadian Forest Fire Danger Rating System. 

\subsection{Cell-based fire growth models}

Cellular automata models that employ a raster-grid of square or hexagonal cells, are widely used to model wildfire spread. Fuel and terrain conditions are usually assumed to be homogeneous within each cell in order to simplify basic fire spread rate calculations. The fire propagates through the grid-cells basis, typically from a cell's center to the center of an adjacent cell. Each ignited cell behaves as an ignition source that is independent of any adjacent burning cells.  To spread the fire from one cell to another, a search mechanism such as an adjacency or spread template is required.

\ignore{Kourtz Regan }\cite{Kourtz1971} developed the first computer simulation model to spatially simulate the growth of a small fire. Their model was based on a heterogeneous and discontinuous fuel-type grid but did not account for the effects of terrain and wind. This deterministic model predicted how long it would take a fire to burn through one square area or cell within a fuel grid when the location of the fire, the starting time, and the grid resolution were known.
Travel times were calculated using fixed rates of spread (based on the fuel type and the spread index for the day) and fixed spread directions from the burning cell. Later,\ignore{ O’Regan et al. in} \citet{ORegan1973} developed a method for using directional rates of spread to predict fire growth. They also rewrote the original model for use on what was then a large computer to simulate fires of up to 15,000 ha. in size.

\cite{ORegan1976}\ignore{ O’Regan et al.} calculated average directional rates of spread using the equation:
\smallskip
\begin{equation}
    ROS(\phi)=\left\{ \begin{array}{cc}
    \dfrac{a\left(1-e^{2}\right)}{1-e\cdot cos\phi}, & 270\text{º}<\phi<90\text{º}, \\
    a\left(1-e^{2}\right), & 90\text{º}\leq\phi\leq270\text{º},
\end{array}\right.
\end{equation}
where $a$ and $b$ are major axes, the quotient $c/a$ is the eccentricity $e$ and $\phi$ is the focal angle. \ignore{Kourtz et al. }\citet{Kourtz1977} then further modified this model to accommodate variation hourly wind conditions. \cite{Todd1999} adapted the\ignore{ Kourtz et al.} \citet{Kourtz1977} model to create an eight-point symmetric fire growth model called Wildfire, which incorporates FBP System spread rates. The features and functionalities of the Wildfire model were assessed and considered during the design of the Prometheus model \citep{Prometheus}. 

\ignore{Boychuk et al. in }\cite{Boychuk2009} developed a stochastic model of fire spread using a lattice Markov chain model, in which they associated probabilistic transition functions with each cell. Each of these cells interacts with its four nearest neighbors and a cell transitions from unburned to burning depending on the state of the neighboring cells. The use of a simplified cellular automata model describing the dynamics of fire spread on a heterogeneous landscape accounting for weather factors (wind speed and direction) as well as the type and density of vegetation was used to successfully model the Spetses Island fire, Greece was after tuning the main parameters of the simulator \citep{Automata2008}.

Recently, a fire growth simulator \citep{Arcaetal} was released with the goal of assisting civil protection and fire management agencies. In order to provide short term forecasts of how the fire will grow, they based their simulator on models of wind field progression. The fire growth model uses the level set technique (see \cite{ghisuetal} and the\ignore{ Rothermel} \cite{Rothermel1972} fire behavior model.) The goal of our paper, in contrast, is to not to support large fire management operations but rather to inform strategic harvest planning and fuel management. We are interested in producing final fire scars that are reasonable. To do that, we need to have a reasonable evolution of the fire, but approximations with respect to minute-by-minute status are acceptable in the interest of computational efficiency because many scars may need to be generated to explore methods for strategic timber harvest scheduling and forest management planning or to validate the use of simpler methods for creating fire
scar scenarios such as, e.g., \cite{ellipses}.

\subsection{Canadian Fuel Behavior Prediction System: FBP}
\begin{figure}[h!]
	\centering
    \includegraphics[scale=0.50]{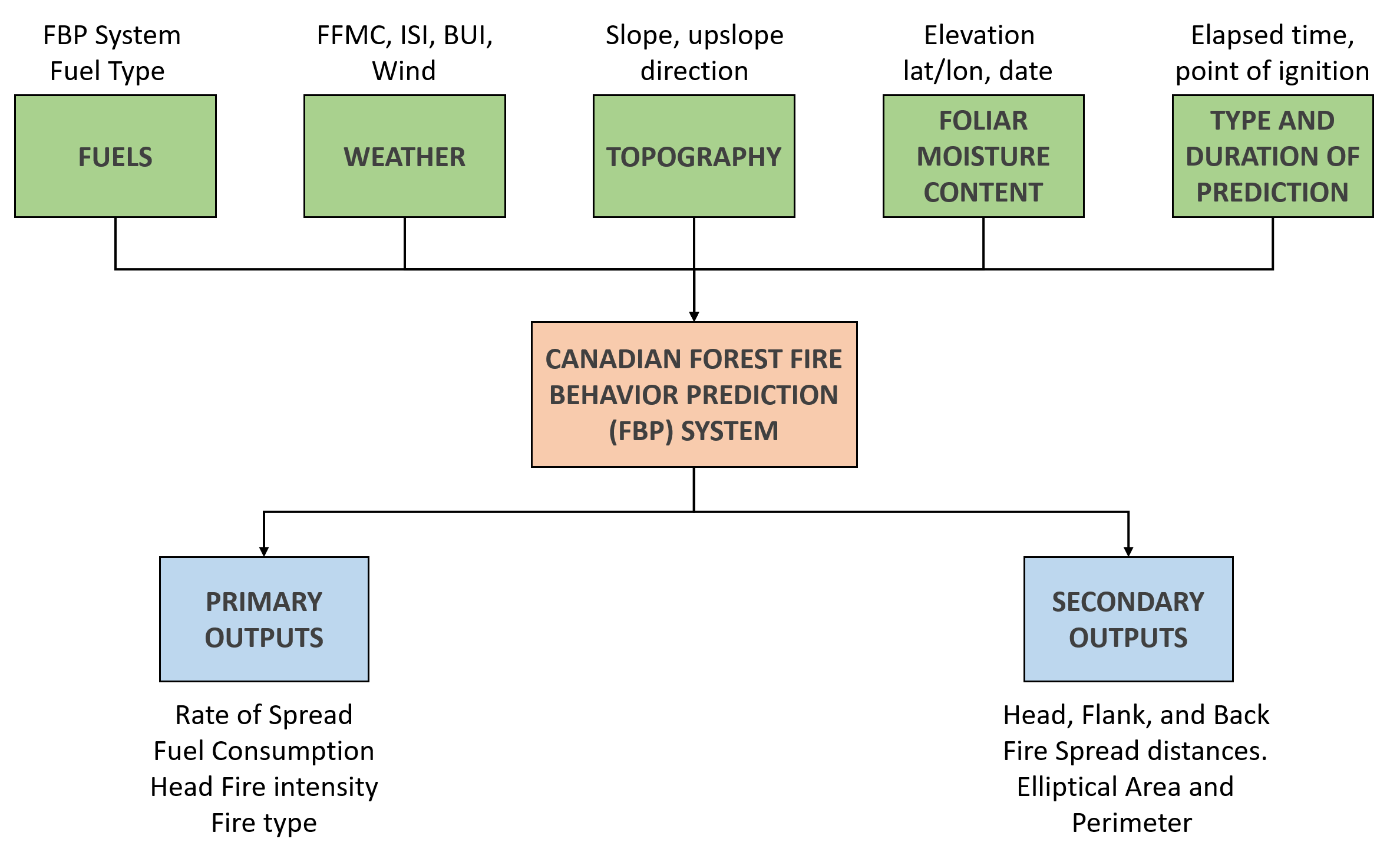}
    \caption{A conceptual diagram of the Canadian Forest Fire Behaviour Prediction (FBP) System. }
    \label{FBPSystem1}
\end{figure}

The Canadian FBP System is a set of empirical models that can be used to predict fire spread rate, fuel consumption and fire intensity in homogeneous fuel types as functions of fuel type, fuel moisture and current weather expressed in terms of the Canadian Forest Fire Weather Index System (FWI) codes and indices \citep{FBP}.
It includes fuel models that are used to classify forest types into 17 fuel types that collectively represent most of the major forest cover types in Canada. Figure \ref{FBPSystem1} is based on a figure in \cite{Prometheus} and shows the main inputs and outputs of the FBP System.

\section{Cell2Fire growth simulator}
\subsection{Description}  

Cell2Fire is an open-source cell-based fire growth simulator developed in Python and C\textbf{++} for use on laptop or desktop computers as well as on High-Performance Computer (HPC) systems. It allows a user to simulate  fire growth across a grid that represents a real forest landscape using fire environment variables such as the fuel type, elevation (topographic/terrain components) of each cell, fire ignition points and weather.

A forest landscape is mapped into a rectangular region comprised of $n$ rows and $m$ columns partitioned into a series of identical area square cells to produce a grid where the cell size depends on the desired spatial resolution that will of course, be influenced by the spatial scale for which  cell attribute data is available. Each cell represents a specific portion of the landscape and has two information layers that pertain to its topographic and fuel characteristics. Those layers define the  characteristics of each independent cell, allowing the simulator to treat them as individual objects that can interact with other cells on the forest landscape to model fire growth. 

Algorithmically, Cell2Fire simulates the growth of fire by tracking the state of all cells as the model progresses through discrete equally-spaced time steps. The status of the fire and all the cells on the landscape are updated at the end of every time step and smaller time steps lead to better fire growth precision.

A cell can be in one of five states: \textit{``Available''}, \textit{``Burning''} , \textit{``Burned''}, \textit{``Harvested''}, or \textit{``Non-Fuel''} where the label \textit{``Available''} indicates that the cell contains a flammable fuel type; \textit{``Burning''} represents that the cell contains an active fire; \textit{``Burned''} indicates that the fire has passed through the cell;  \textit{``Non-Fuel''} is a non-flammable fuel type such as rivers, lakes, or rocky. The state \textit{``Harvest''} is provided so that the simulator can be embedded in a strategic harvest planning system. The harvest planning module would be responsible for labeling the cells that are harvested and provide the appropriate post-harvest fuel type.

Once an ignition point has been specified, the fire is ignited. During each time step the fire may spread along the axes emanating from the centre of each burning cell to its neighboring cells. The predicted FBP system Head Rate of Spread (HROS), Flank Rate of Spread (FROS) and Back Rate of Spread (BROS) are used to model elliptical fire growth within each cell with the focus of the ellipse at or near the center of the cell (see this process described in more detail in \ref{sec:ROS}).  The geometry of the ellipse is then used to predict the fire spread rates along the axes emanating from the centre of each cell.

\begin{figure}[h!]
	\centering
    \includegraphics[scale=0.80]{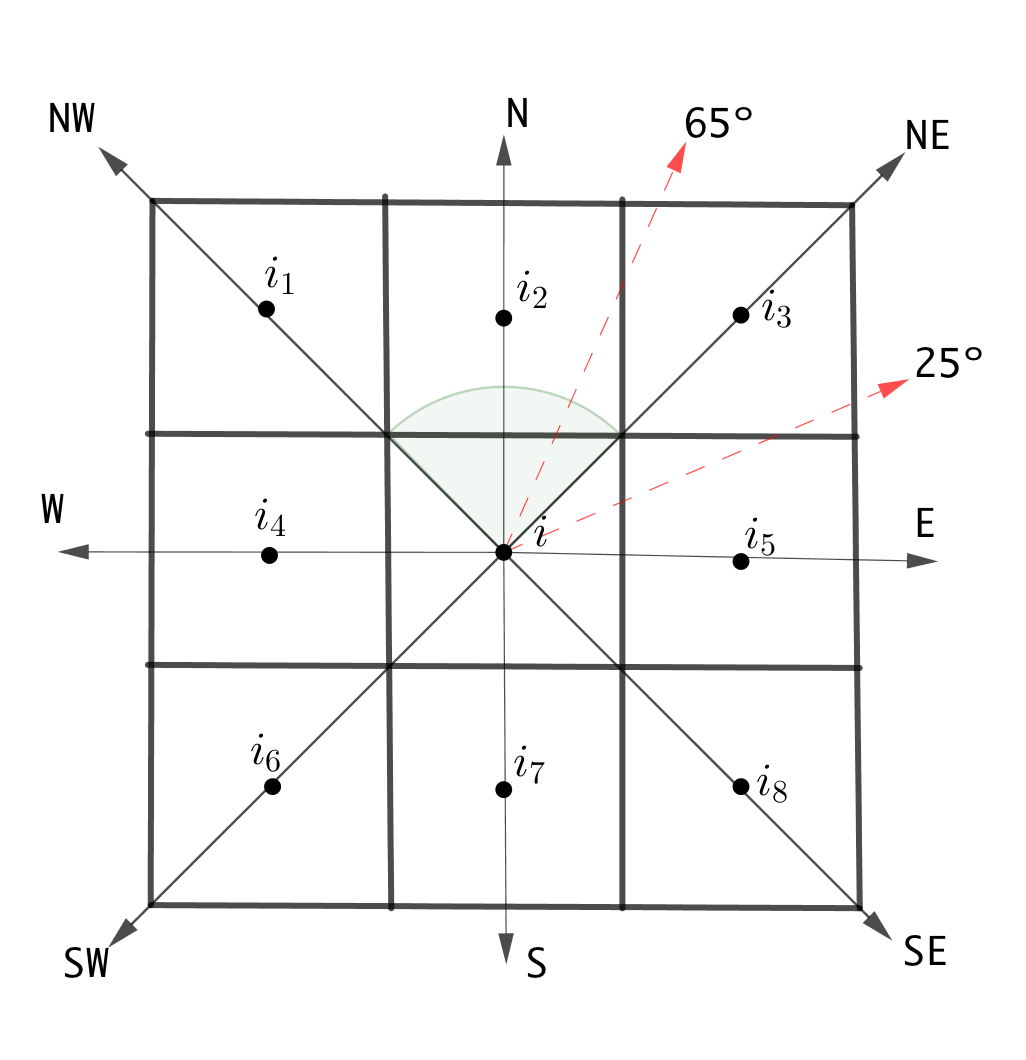}
    \caption{Neighbors cells.}
    \label{NCells}
\end{figure}

A signal/message is then initiated to any adjacent cells whose center is reached by the fire. In the present implementation, it is assumed that each cell has at most 8 adjacent cells (see Figure \ref{NCells}) because the grid is assumed to be rectangular. 
These are the only neighbors considered because the simulation 
time step is assumed to be small enough to ensure that the fire can not spread beyond adjacent cells in one time step. 
For the examples we tested using a 100m by 100m cell size, simulation time steps below one minute resulted in almost no change in final fire scar compared to one minute. Shorter
simulation time steps result in longer run times for the simulation.

Note that there is also an approximation because the fire enters a cell a from its neighbor and the neighbor's cell
characteristics are used to model fire spread within that cell until the fire reaches the center of the destination cell. At that point,the characteristics of the destination cell take over.

When the fire spreads from the edge of a cell and reaches its center, the cell receives a message. When the cell receives a message it calculates what its ROS based on its characteristics and the current weather and if that ROS is greater than a user-specified parameter (which was zero in the experiments reported here), its state is labeled as \textit{``Burning''}. At this point, the main Rate of Spread values are calculated by the FBP System module and fire progress begins to be calculated along with its available axes. Predicted ROS values for each axis are based on the assumption that fire spreads in the shape of an ellipse in each cell and the geometry of the ellipse and its orientation are used to predict spread rates along the axes emanating from the cell as described later in Section \ref{EllipseFitting}.

The fire's progress is updated at each fire time period by examining the state of all \textit{``Burning''} cells. Once no adjacent cells are available or a burn-out criterion (See assumption (A5) in \ref{Burn-out}) has been satisfied the cell changes its status to \textit{``Burned''} and is omitted from the simulation in further steps.

This process is repeated until a fire ending event state is reached, i.e., the total number of fire burning periods or hours have passed or there are no more cells available to burn.

Three different sources of uncertainty can be included in our model to account for the most important perturbations that can significantly affect the fire growth dynamics: (1) ignition point(s) selected via a user-defined spatial probability distribution or else simply with equal probability for each cell, (2) a perturbation level the stochastic aspects of the basic FBP predicted Rates of Spread via its coefficient of variation, accounting for its inherent approximation error, allowing the user to obtain different fire scars taking into account uncertainty in the fire spread rates, and (3) a set of user-generated weather stream files (fire weather scenarios that can occur with specified probabilities can be provided to Cell2Fire, performing a series of simulations with different weather scenarios thereby allowing the user to study the behavior of the fire under different weather conditions for a specified forest/instance.  

A series of relevant outputs are generated depending on the user needs: fire evolution maps (scars) at different time-precision resolutions, burn grids that indicate the final status of each cell for statistical analysis as well as fire probability maps if there are probabilistic inputs, and a simulation log that documents all the interactions between cells and statistics of the final state of the forest. 

\begin{figure}[h!]
	\centering
    \includegraphics[scale=0.4]{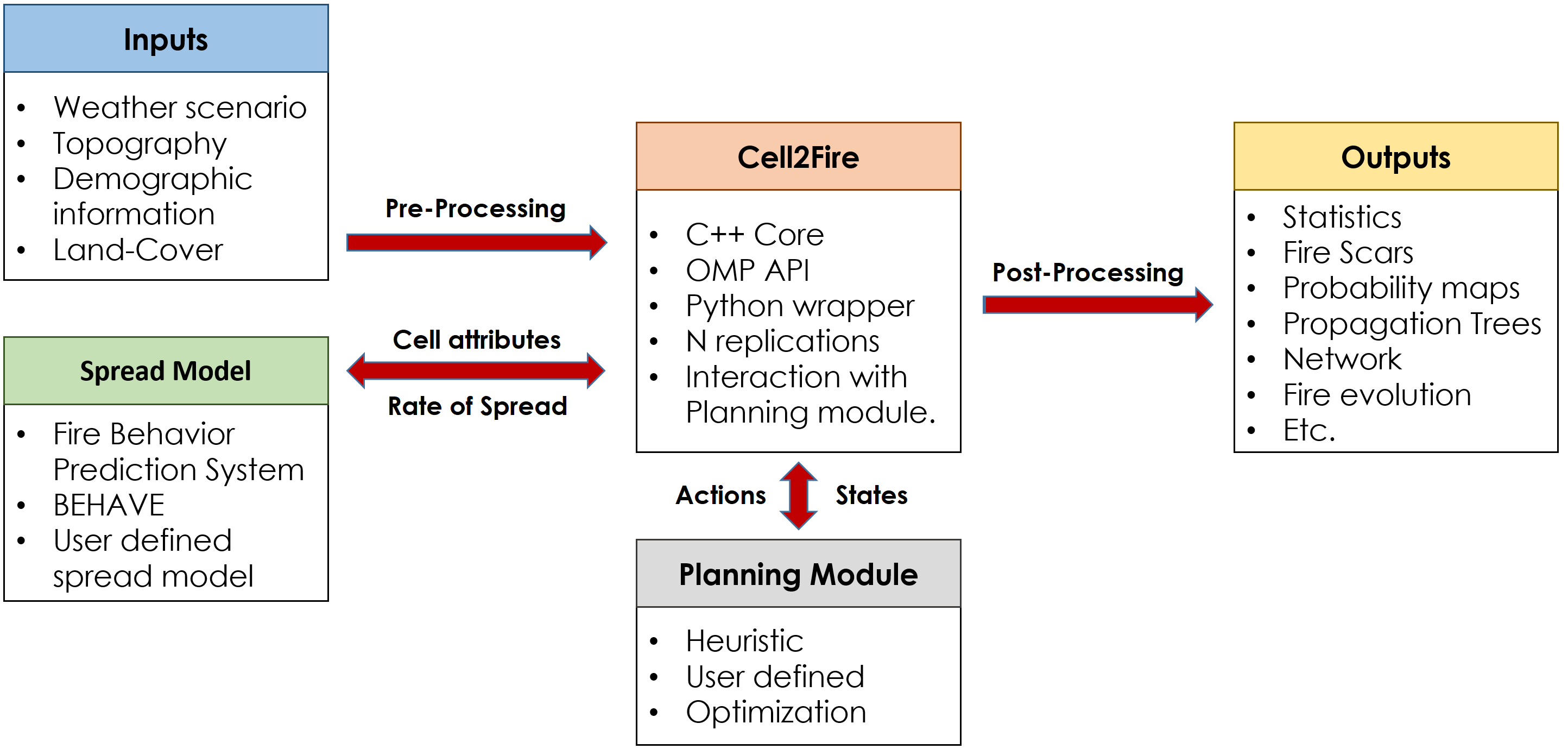}
    \caption{Simulation framework. (1) Raw data is pre-processed into Cell2Fire's format, (2) Cell2Fire calls an independent fire spread model (e.g., FBP), performing the simulations including harvesting plans provided by the user (if needed), and (3) finally, outputs are generated and returned to the user.}
\end{figure}

\subsection{Modelling assumptions}
Based on the previous description description of our model we can summarize the main simplifying assumptions upon which Cell2Fire is based as follows:

\begin{itemize}
\item[(A1)] The growth of the fire depends on the ``Rate of Spread (\textit{ROS})'' of the fire from a burning cell towards its neighboring cells. We assume that a cell is ignited when the fire reaches its center and conditions for burning are met (see A5). Each cell has at most 8 neighbors.

\item[(A2)] The Rates of Spread along the 4 principal orthogonal axis (which are functions of the weather, fuel moisture conditions and characteristics of each cell) are calculated using the Canadian FBP System. Note that a the main axis of the ellipse that burns in each cell is aligned in the direction that the wind is blowing (the HROS direction) for that time step and the BROS is aligned in the opposite direction.  The FROS is perpendicular to the wind direction (the two flanks). Note that other fire spread models could be used in lieu of the Canadian FBP system, a point we discuss in our Conclusions section. 

\item[(A3)] Each cell that burns serves as a new source of fire. Fire spread occurs at the cellular level and cell size depends on the spatial resolution and corresponding data availability.

\item[(A4)] The effect of fire suppression action is not modelled. It is beyond the scope of this paper and will be addressed in a future research project.

\item[(A5)] \label{Burn-out} There are two sets of conditions for modeling the termination of fire growth  (a fire ending event) in Cell2Fire: the cellular level and general fire evolution conditions. 
At a cellular level, each cell becomes unavailable (burned) for future fire dynamics if (i) the ROS along the main available axes is less than some empirical threshold $\delta > 0$, (ii) the cell does not have any adjacent cells that are available to burn, (iii) the residual fuel available in a cell is not sufficient to support fire spread (implicit in the FBP system), or (iv) a user-defined head fire intensity (HFI) threshold is provided and the HFI is below that threshold.

Regarding the general fire dynamic, the total duration of the simulated wildfire event is determined by (1) maximum number of hours of burning per day --- a season-dependent constant \cite{BurnP3}, or drawn from a probability distribution -- and (2) the total fuel remaining in the forest (available cells).

\end{itemize}

\subsection{Cells}
Cells that contain the information concerning the structure of the forest as well as the past and current state of the simulation are the main processing units in Cell2Fire. The main fields that define the state of a cell $i \in \mathcal{N}$ are the following:
\begin{itemize}
	\item[i)] Fuel type: following the classification criteria of the FBP System \citep{FBP}, each cell is assigned a specific fuel type (e.g., conifer, grass, non-fuel) represented by an FBP code. This will be used for selecting the specific fire spread models and coefficients defined by the FBP System in order to predict the rate of rate of spread (ROS) in that cell. 
	
    \item[ii)] Slope: the slope \% in terms of the vertical rise over the horizontal run and adjacent cells, and the upslope direction (radians) have a significant impact on the predicted ROS. 
    \item[iii)] Elevation: altitude in meters of the current cell with respect to the sea level.

    \item[iv)] Location: latitude and longitude coordinates are provided with the instance.
    \item[v)] Status: cells are classified into five different states during the simulation time steps. Cells that have not been harvested/burned and represent valid fuel types are labeled as \textit{``Available''} cells. Cells that are actively burning are classified as \textit{``Burning''} cells, updating the fire progress along each of its 8 axes every time-step. 
    When a burned-out condition (see Section 3.2) is reached, burning cells are turned into \textit{``Burned''} cells, meaning that they can no longer contribute to the propagation  of the fire across the forest and can be omitted in further simulation steps. Finally, a cell can be labeled as \textit{``Harvested''} or \textit{``Non-Fuel''} and therefore, not available for the fire spread.
    \end{itemize}

In addition, each cell contains a series of secondary parameters that allow Cell2Fire to track the evolution of the fire within the forest and change the state of the current simulation run.  The Fire dynamics group includes the fire ignition date and time of each cell, the fire's progress along each axis, and the effective $ROS(t,\theta)$ values (per period and axis angle). The average age of trees inside the cell, approximate volume of wood/products available, ID label, perimeter, area, adjacent cells list, and distance to adjacent cells centers can be found in the {\em Demographic} category. 


Due to this independent structure, cells can be treated as individual units allowing an efficient parallel computing approach for each iteration. We can therefore update their status and generate the relevant fire messages to model the fire dynamics of each burning cell in parallel using independent threads and update the global status of the forest at the end of each fire time period and thereby obtain significant improvements in execution times (from hours in serial mode to minutes in parallel mode) when dealing with large fires that have many simultaneously active cells.

\subsection{Fire propagation dynamics}
As described above, the fire growth model is simple but powerful: every time a cell is ignited by an adjacent cell it acts as a new source of potential ignition for neighboring cells in the forest, updating the progress of the fire for each available axis (center-to-center directions).

Following an object-oriented programming paradigm in Python and C++, a series of classes for the most relevant components of the problem were developed: Forest, Cells, Weather, Ignitions, FBP System methods and Input/Output formatting. Then, the main program instantiates the different objects and applies the pertinent methods required to simulate fire growth.
Once the fire instance data has been read and the forest has been initialized inside the simulator engine, the main simulation steps are as follows:

\begin{enumerate}

\item [i)] Relevant fire parameters are calculated by performing calls to the FBP System module to determine the Rate of Spread (ROS) for each available fire spread axis of the burning cell: based on fuel characteristics, topography, and weather different $ROS(t, \theta)$ are obtained where $t$ is the current fire time period and $\theta$ is the angle with respect to the center of the adjacent cells (0\degree = East, increasing counter-clockwise) based on the procedure described in Section \ref{EllipseFitting}. Following a discrete time simulation approach, the internal simulator clock advances one unit of time --- a user-input precision parameter --- and the fire's progress is updated for each axis. \\


\item [ii)] Fire spread between cells is modeled by using a sending/receiving message approach (which enables parallelization) based on the computed ROS along each axis. If the fire reaches the center of a cell during the simulated time step, a message is sent. Checking environmental and cells characteristics, a the cell begins burning (or not). This is the core of the simulator and thus, the critical performance bottleneck that comes into play when simulating fire spread across large lanscapes. However, we designed it to maximizing the parallel performance of the code, obtaining a large percentage of naturally parallelizable code, representing around 80\% of its structure.\\

\item [iii)] The previous steps are repeated until some specified ending criterion is satisfied: e.g., the maximum number of weather periods, the maximum simulation time, and/or some fire ending event condition. Statistics regarding the status of the forest as well as plots and other outputs of the fire scar evolution are produced.

\end{enumerate}

\begin{figure}[h!]
	\centering
    \includegraphics[scale=0.65]{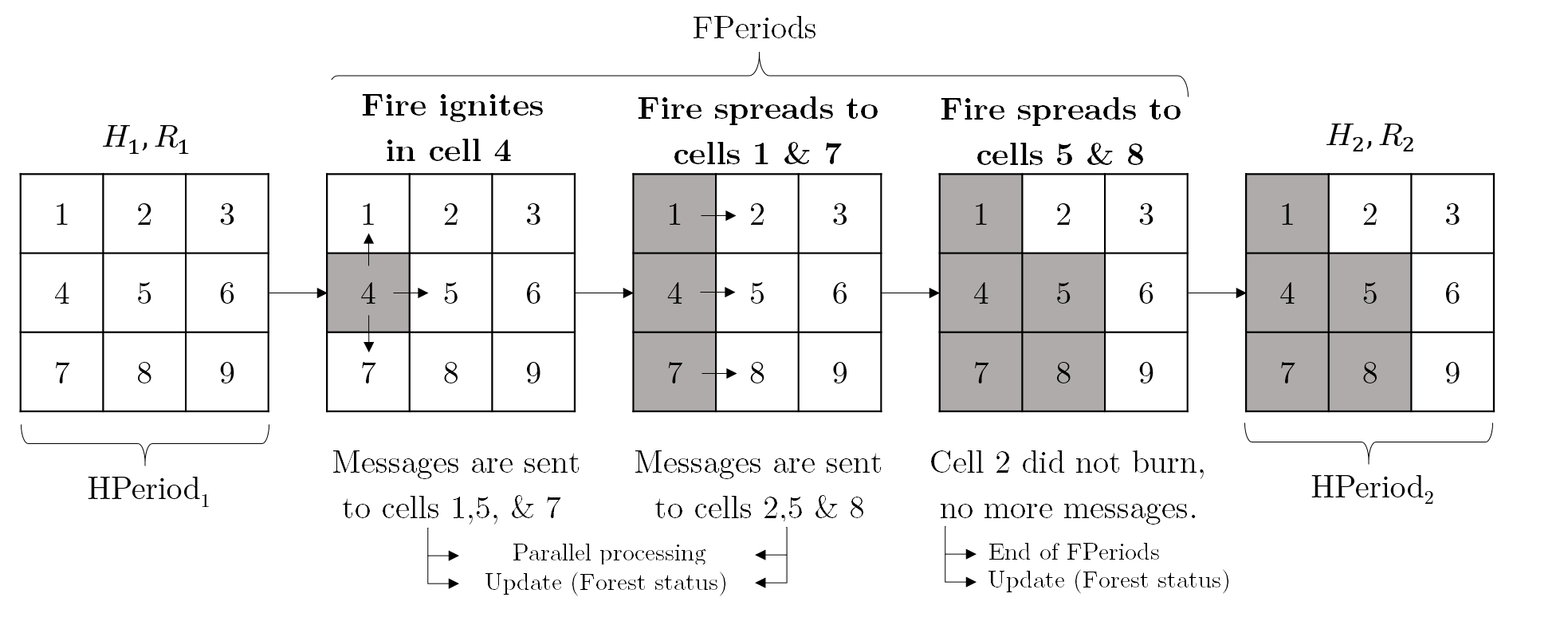}
    \caption{Simulation scheme: Send/Receive messages structure allows natural parallelization. Messages are sent when the fire reaches the center of an available adjacent cell.}
    \label{Sim}
\end{figure}

Figure \ref{Sim} illustrates a forest with 9 cells in which a fire ignites in cell 4, after a harvesting period (e.g., years) from which it can spread to other cells or forest stands for two more fire spread time periods (minutes). If no messages are sent to neighboring cells based on the current environmental conditions (burned-out conditions) or the maximum simulation time for the current fire has stopped growing, go to the time the next fire ignites (randomly generated or user-provided) or stop the simulation. A general pseudo-code of the simulation steps can be seen in Algorithm \ref{SimSteps}. 

We use the following notation to describe the main steps of the simulation Algorithm \ref{SimSteps}:
\begin{itemize}

\item[-] $\mathcal{N}$ denotes the set of cells in the forest and $i\in \mathcal{N}$ represents cell $i\in \mathcal{N}$ .
\item[-] $Adj\left(i\right)$ denotes the set of available  cells that re adjacent to cell $i$ (at most 8).

\item[-] $BurningCells$ is the set of actively burning cells in the simulation process.
\item[-] $i\rightarrow_{t} j$ indicates that cell $i$ ``sends a message'' to cell $j$ at time $t$

\end{itemize}

\begin{algorithm}[h!]
\caption{Cell2Fire Pseudo-code}\label{SimSteps}
\begin{algorithmic}[1]
\Procedure{Sim}{$ForestData,FTypes,Ignitions,Weather,TMax,Options$}
	\State $\textit{\textbf{Step 0}: \textbf{Initialize Cell2Fire}} $
		\State \hspace{1.0cm} \textit{Parse inputs, options, read data, initialize objects}
        \State \hspace{1.0cm} $FPeriod \leftarrow 0, nsim \leftarrow 0$
	\State $\textit{\textbf{Step 1}: \textbf{Ignition}}$
      	\State \hspace{1.0cm} \textit{$ic \leftarrow $ Choice(Ignitions)}
		\State \hspace{1.0cm} $BurningCells \cup \lbrace ic \rbrace $ 
		\State \hspace{1.0cm} $FPeriod \leftarrow 1$
    \State $\textit{\textbf{Step 2}: \textbf{Fire Dynamics (Send-Receive)}}$
		\State \hspace{1.0cm} \textit{Let} $i \in BurningCells$, \textit{\textbf{if}} $i \rightarrow_{FPeriod} j$, \textit{where} $j \in Adj(i):$         
        \State \hspace{2.0cm} $BurningCells \cup \lbrace j \rbrace$
        \State \hspace{1.0cm} $FPeriod \leftarrow FPeriod + 1$, \textit{Update Forest, Weather}
		\State \hspace{1.0cm} \textbf{Repeat, until} $FPeriod < TMax$
	\State $\textit{\textbf{Step 3}: \textbf{Results and Outputs generation}}$
		\State \hspace{1.0cm} \textit{Generate Grids, Statistics, Output plots}
\EndProcedure
\end{algorithmic}
\end{algorithm}

\subsection{Main Inputs}
As is the case with other state-of-the-art fire growth simulators, the Cell2Fire model requires a number of inputs including a minimum set of data layers that define an instance/forest to start the simulations. The relevant inputs needed to simulate the growth of a fire are as follows:
\begin{itemize}
  \item[i)] Forest raster data: ASCII grid forest files that specify the number of cells in the forest, their geographical coordinates and information concerning each cell including as its fuel type, elevation, slope (\% and azimuth), degree of curing.
Files can be in either .csv or .asc format. If topographic data is not fully available, default input dummy values for parameters such as elevation or slope can be provided by the user (null by default).

  \item[ii)] Fuel type dictionary: Fuel type codes and descriptions that match the Canadian FBP System fuel types (Table \ref{FBPT}) are included inside a .csv file. Future implementations will allow the use of a custom dictionary file that includes fuel types not currently included in the set of Canadian FBP fuel types.

\bigskip  
  \begin{table}[h!]
  \centering 
  \label{TableFTypes}
  \begin{adjustbox}{max width=0.7\textwidth}
\begin{tabular}{@{}cccc@{}}
\toprule
\textbf{grid\_value} & \textbf{export\_value} & \textbf{descriptive\_name}    & \textbf{fuel\_type} \\ \midrule
1                    & 1                      & Spruce-Lichen Woodland        & C-1                      \\
2                    & 2                      & Boreal Spruce                 & C-2          \\
3                    & 3                      & Mature Jack or Lodgepole Pine & C-3          \\ \bottomrule
\end{tabular}
\end{adjustbox}
\caption{FBP fuel type dictionary sample. The grid\_value field refers to the encoding of the forest inside the ASCII files and the fuel\_type column contains its translation into the FBP code.}
\label{FBPT}
\end{table}
  
  \item[iii)] Ignition points: An optional file that that specifies the cell(s) in which fires are to be ignited during the simulation, paired with their corresponding ignition time periods.
  
    \item[iv)] Weather stream: Hourly weather records for one or more fire weather stations located near the area of interest that include the date-time, precipitation, temperature, wind speed/direction, relative humidity, scenario ID, as well as the daily fire danger rating codes and indices (FFMC, DMC, DC, ISI, BUI, and FWI) of the Canadian Forest Fire Danger Rating System \citep{CFFDRS}, registered by the stations and equations inside the FBP System module (Table \ref{WeatherT}). Data from the nearest weather station is used for each cell.

\bigskip  
\begin{table}[h!]
\centering
\label{TableWeather}
\begin{adjustbox}{max width=0.8\textwidth}
\begin{tabular}{@{}cccccccc@{}}
\toprule
\textbf{Scenario} & \textbf{datetime} & \textbf{APCP {[}mm{]}} & \textbf{TMP {[}C{}\degree{]}} & \textbf{RH {[}\%{]}} & \textbf{WS {[}m/s{]}} & \textbf{WD{[}\degree{]}} &  \\ \midrule
JCB               & 2001-10-16 13:00  & 0.0                    & 17.7                                       & 20                   & 21                    & 225                                                        \\
JCB               & 2001-10-16 14:00  & 0.6                    & 16.9                                       & 18                   & 25                    & 205                                      \\
JCB               & 2001-10-16 15:00  & 1.2                    & 16.1                                       & 20                   & 27                    & 190                                                \\
JCB               & 2001-10-16 16:00  & 10.0                    & 15.8                                       & 20                   & 37                    & 232                                                \\
JCB               & 2001-10-16 17:00  & 5.3                    & 13.9                                       & 25                   & 43                    & 225                                                \\
JCB               & 2001-10-16 18:00  & 2.1                    & 12.1                                       & 35                   & 45                    & 222                                                \\
JCB               & 2001-10-16 19:00  & 0.9                    & 10.6                                       & 41                   & 46                    & 241                                                \\
JCB               & 2001-10-16 20:00  & 0.0                    & 11.3                                       & 39                   & 18                    & 248                                                \\ \bottomrule
\end{tabular}
\end{adjustbox}
\caption{Extract of an hourly weather stream file. Average precipitation (APCP), temperature (TMP), relative humidity (RH), wind speed (WS) and wind direction (WD).}
\label{WeatherT}
\end{table}

\end{itemize} 

Sample files are included with the distribution of Cell2Fire for the publicly available Dogrib instance (\url{http://www.firegrowthmodel.ca/prometheus/software_e.php}) as well as simple generated testing instances. 

Besides the main input files, a set of options and user-provided parameters for exploiting all the flexibility of the simulation engine including tuning options (see Section \ref{EllipseFitting}) are available as secondary inputs when running Cell2Fire.
   
\subsection{Main Outputs}
Once a simulation run has been completed, the following outputs are available:
\begin{itemize}
  \item[i)] Burn-Grids: Files in which 1s indicate burned cells and 0s indicate those that are available to burn. Useful for statistical comparisons with other simulators as well as to generate burn probability maps.
 
  \item[ii)] Plots: Initial forest state, fire scar evolution, and message sending/receiving can be visualized by a series of plots (Figure \ref{InitStateF}) generated after the simulation run has been completed.
  
\begin{figure}[h!]
 	 \centering
     \includegraphics[scale=0.65]{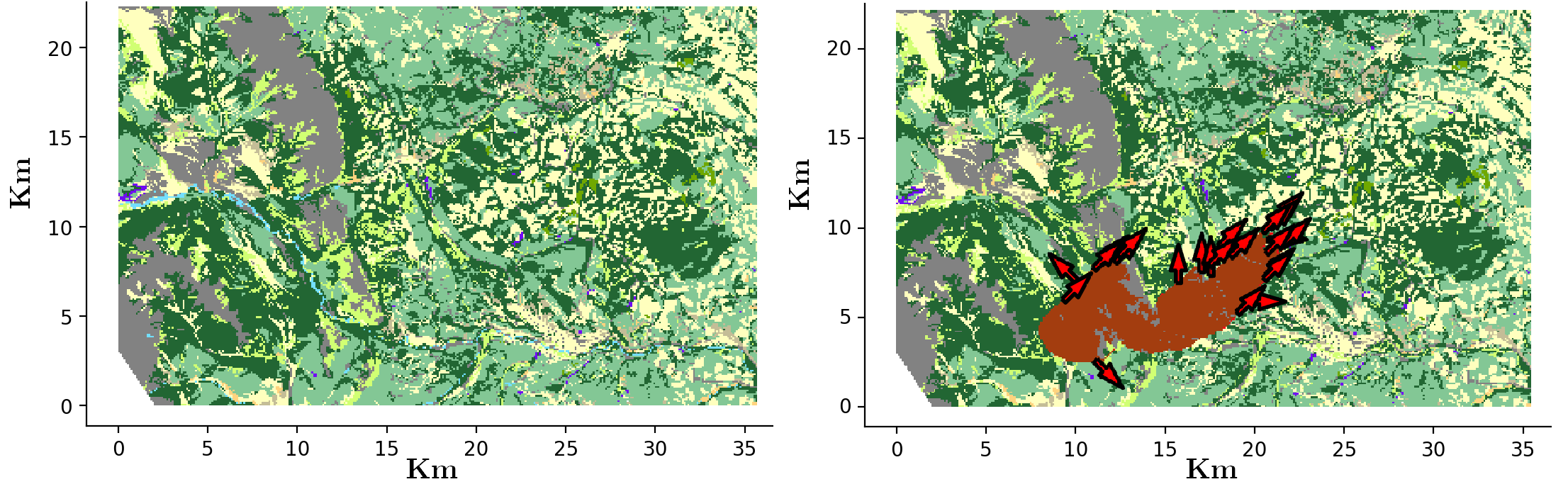}
     \caption{Initial forest state (left side) plot example for Dogrib instance. Each color represents a cell with a specified fuel type, encoded according to the FBP system. Fire scar evolution example (right side) including burning cells (orange) and new fire messages directions.}
     \label{InitStateF}
 \end{figure}

  \item[iii)] Statistics: Final status of the forest with relevant information regarding the messages, fire ignition dates, fire evolution (distance per axis and ROS values), average number of burned/available cells, the average percentage of the forest available for all the tested scenarios (including multiple ignition points, weathers).
  
\end{itemize}

\subsection{Rate of Spread from cell to cell} 
\label{sec:ROS}
\label{EllipseFitting}

We use the approach proposed by the Canadian FBP System \cite{Prometheus} where an elliptical distribution scheme is applied. This method does not require the ignition point or the point of propagation to coincide with either of the two foci of the ellipse, although the authors indicate that ``\textit{small differences between the point of ignition and the focus of the ellipse do not change the results}'', i.e., the elliptical propagation shape/size.  
\begin{itemize}
    \item[-] An elliptical fire has a HROS (head), a BROS (backing) and a FROS (flank). Let $a$, $b$, and $e$ be the semi-major, semi-minor, and eccentricity of the ellipse, respectively.
    \item[-] The FBP system predicts the HROS, the BROS and the length to breadth ratio (LB) which is 2a/2b or a/b.
    \item[-] During the first time interval $t$, the fire will spread from its ignition point to the center of the ellipse and then from the center of the ellipse to the farthest edge of it.
    \item[-] At time $t$, we have:
    \begin{eqnarray}
        a &=& \dfrac{HROS + BROS}{2} \times t,\\
        b &=& \dfrac{2 \times FROS}{2} \times t,
    \end{eqnarray}
    to expand the ellipse generated by the propagation of fire at time $t$ on the two main axes.
    \item[-] Noting that $LB = \dfrac{a}{b}$ we have $FROS = \dfrac{HROS + BROS}{2 LB}$. Therefore, the eccentricity is: 
    \begin{eqnarray}
        e = \sqrt{ 1- \left( \dfrac{{(FROS \times t)}^{2}}{{ {\dfrac{(HROS + BROS) \times t}{2}}}^{2}} \right)   }
    \end{eqnarray}
\end{itemize} 
Using these equations and the procedure described in \citet{Prometheus}, we can estimate the Rate of Spread from the center of a cell to the center of any adjacent cell as in equation (1).

\begin{figure}[h!]
	\centering
    \includegraphics[scale=0.3]{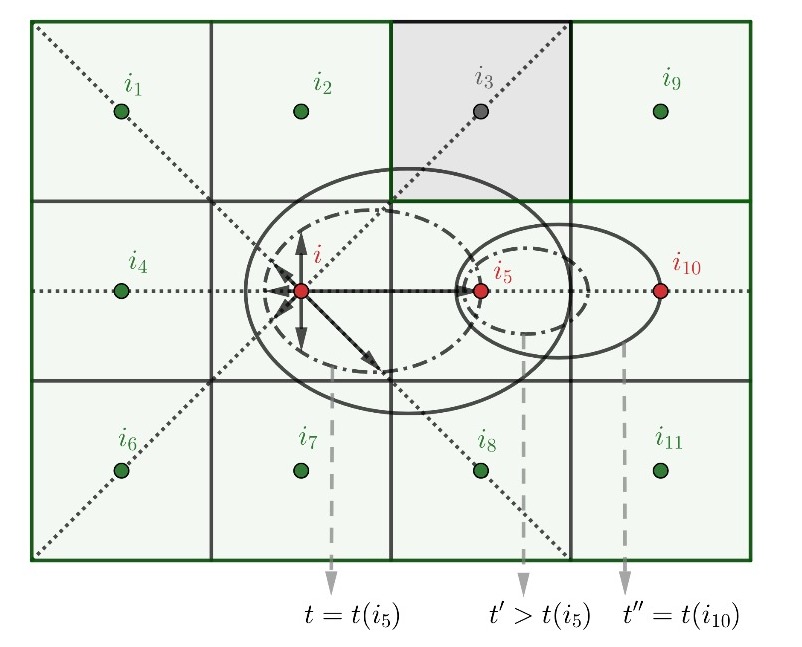}
    \caption{Elliptical Rate of Spread distribution scheme using the ellipses defined by the Canadian FBP System. At any time $t$, the backfire will be $BROS \times t$ behind the point of ignition and the head fire will be $HROS \times t$ ahead of the point of ignition of the fire, expanding the ellipse. Then, if the fire emitted by cell $i$ reaches the center of an adjacent cell $j$ at time $t(j)$, a new ellipse is generated at time $t' > t(j)$, calculating the corresponding rate of spreads, and so on.}
    \label{CellularHuygens}
\end{figure}

\subsection{Computational details} 
     We chose C++ for the parallel implementation because the objects we created in Python are naturally mapped to C++ objects, obtaining at least 15-20x average speedups depending on the forest characteristics. As previously discussed, our algorithm contains 3 sections during each time step: (1) checking for new lightning fire ignitions (igniting), (2) updating the intensity of already-burned cells and analyzing newly burned cells (sending messages), and (3) marking newly burned cells as burning (receiving messages). 
     
Analyzing Cell2Fire running times, the ignition stage is very quick, with most simulations only igniting a single time at the first time step of the simulation. The sending-messages stage updates the fire progress in every burning cell. Because we can have a large number of cells burning at once and there are no direct dependencies on neighboring cells, this part is easily parallelizable. Each cell, in addition to updating its current status, can also ``send a burning message'' to an adjacent cell. In the receiving messages stage, we analyze the ``burn messages'' sent to non-burning cells and mark them as burned if the conditions are met. This part is also potentially parallelizable, but because the number of newly burned cells at a single time-step is dwarfed by the number of currently burning cells, we found that a speedup here is of lower priority.

Due to the easily parallelizable structure of our code, the most suitable approach for parallelizing its execution consists of a shared-memory approach using the well-known OpenMP API \citep{OpenMP}. This is an advantage of Cell2Fire since the code will be also optimized for its execution in daily-use computers, without needing a multi-node architecture for exploiting its parallelism. Using it, we were able to obtain an average of 15\%-20\% extra performance for the parallel region.

\section{Results and Discussion \label{sec:experiments}} 
In this section, we compare the predicted burn scars produced Cell2Fire and Prometheus for several instances created for this purpose as described in Section~\ref{sec:instances}. We did not compare either simulator with the realized fire scars in our study because it is difficult to determine the extent to which the final shape of real fires that were fought was or was not influenced by suppression action.  The simulator is fast and exhibits good parallel speed as described in the Appendix.

\subsection{Comparison Methodology}

The comparison methodology we used consists of measuring the difference between fire scars generated by Prometheus and Cell2Fire simulations using three metrics. The simplest and most widely used full-reference quality metric is the mean squared error (MSE), which objectively quantifies the strength of the error signal. However, two distorted images with the same MSE may have very different types of errors, some of which are much more visible than others. Thus, as we are interested not only in the difference of the marks but also in structural information, we use a measure of similarity suggested in \citet{SSIM} and denoted by SSIM (structural similarity index). Finally, we also include the Frobenius norm of the difference between two scars $X$, $Y$, $\delta_{norm} = ||X-Y||$.
\medskip{}

First, we compare the evolution of Cell2Fire and Prometheus fire scars on a period-to-period basis (where a period represents one hour) in order to measure the differences in the fire propagation. We denote by $PromGrid^{t}$ a 0-1 Matrix at time $t$, which represents the fire scar obtained with Prometheus, where $PromGrid^{t}_{ij}$ is equal to 1 if the cell $\left(i,j\right)$ was burned and 0 otherwise. Analogously, we define the fire scar obtained by Cell2Fire as $Cell2Grid^t$.

Below, $\mu_{X^t}$, $\mu_{Y^t}$, $\sigma_{X^t}$, $\sigma_{Y^t}$ and  $\sigma_{X^t Y^t}$ represents the means, standard deviations, and cross-covariance for scars $X^t$ and $Y^t$ respectively, and $C_1$, $C_2$, and $C_3$, are internal parameters of the metric \cite{SSIM}. The methodology is as follows:

\begin{enumerate}
\item Choose a ignition point for each instance and run Prometheus for 
$T$ time periods. Thus, we obtain $PromGrid^{t}$, $t=1,...,T$
(0-1 Cell Matrices).

\item Choose the same ignition point as above and then run Cell2Fire for
$T$ time periods from this point. Outputs: $Cell2Grid^{t}$, $t=1,...,T.$
\item Set $X^{t}=PromGrid^{t}$ and $Y^{t}=Cell2Grid^{t}$ and calculate
for all $t$:
\begin{enumerate}
	\item Mean Squared Error:
\[
MSE\left(X^{t},Y^{t}\right)=\dfrac{1}{nm}\sum_{i=1}^{n}\sum_{j=1}^{m}\left|X_{ij}^{t}-Y_{ij}^{t}\right|^{2},
\]
to measure average of the squares of the pixel differences of the fire scars, and
	\item Structural similarity Measure:
\[
SSIM\left(X^{t},Y^{t}\right)=\dfrac{\left(2\mu_{X^{t}}\mu_{Y^{t}}+C_{1}\right)\left(2\sigma_{X^{t}Y^{t}}+C_{2}\right)}{\left(\mu_{X^{t}}^{2}+\mu_{Y^{t}}^{2}+C_{1}\right)\left(\mu_{X^{t}}^{2}+\mu_{Y^{t}}^{2}+C_{2}\right)},
\]
to measure the change in structural information between the fire scars obtained from the two simulators: Cell2Fire and Prometheus.
\end{enumerate}

\item Measures analysis: MSE and SSIM evolution for all time $t$, $\delta_{norm}$ for the final scar.

\end{enumerate}

\subsection{Instances \label{sec:instances}}
We used three sets of fire instances to compare the performance of Cell2Fire with Prometheus: (1) Sub-instances set, (2) British Columbia province real landscapes with simulated wildfires and (3) Case study: Dogrib fire landscape. 

\subsubsection{Sub-instances}

We used portions of the Dogrib landscape data (see Section~\ref{sec:dogrib}), we generate two sub-instances that we label Sub-1 and Sub-2 with a cell resolution of $100 \times 100$ meters. The first one represents a sub-forest from the Dogrib landscape that is $20 \times 20 $ cells (400 ha.) and the second a $40 \times 40$ cell (1600 ha.) instance. Both consist of heterogeneous landscapes that include different fuel types as well as non-flammable cells (such as mountains or rivers). An ignition point was selected for each instance as a starting point for the fire simulation. Three weather stream files: Weather-1, Weather-2, and Weather-3 of 6, 14 and 22 hours respectively, were used as main inputs: the first one contains data for the 6 hours during which the real Dogrib fire made a run (extreme weather conditions) while the second and third contain additional meteorological measurements from the same day of the fire, before and after that spread event, thus, being an extension of the Dogrib's weather scenario.

After the ignition point was fixed for both instances, we proceeded to run the simulation in Prometheus and the deterministic version of Cell2Fire, generated the hourly fire scars and calculated the performance indicators (1-MSE and SSIM).

\begin{figure}[h!]
	\centering
    \includegraphics[scale=0.72]{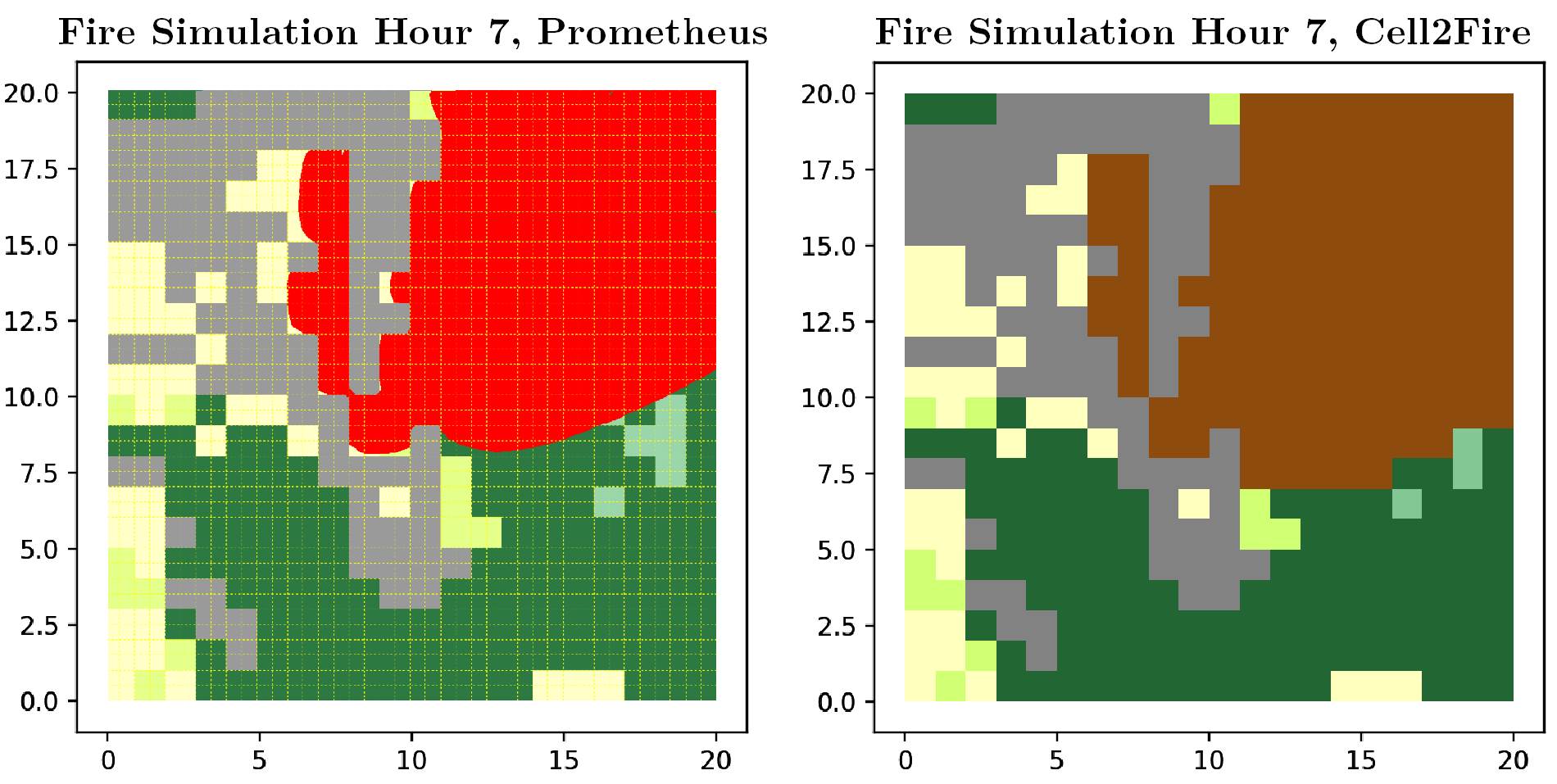}
    \caption{Sub-Instance 2 Fire evolution visualization in Prometheus and Cell2Fire including a heterogeneous landscape with non-flammable cells (mountains, gray cells) and different fuel-types (green and yellow cells).}
    \label{sub2}
\end{figure}   

Based on the results shown in Table~\ref{detresults} and Figure \ref{sub2} we can see that Cell2Fire produced results that are similar with respect to the hourly fire growth and final fire scar to those produced by Prometheus. When testing the 6 critical hours of the Dogrib fire, the level of difference (MSE) is less than 5\% for the first sub-instance and less than 6\% for the sub-instance 2. For the full day simulation, a slight increment of the differences is registered, reaching average MSE levels values close to 6\% and 7 \% for Sub-2 and Dogrib instance respectively.

On the other hand, it is possible to observe how the structural similarity measurement has a decreasing pattern (see Table~\ref{detresults}), a situation that is even more explicit when comparing the 22 hour fire weather stream. This indicates that the fire growth predicted by Cell2Fire differs from the one predicted by Prometheus due to: (1) as expected, the approximation of the elliptical model into an adapted cellular-automata approach implies a different fire dynamic, increasing the differences on every time step when comparing with the wavefront model and (2) differences in the calculations/approximations of the effective $ROS(t,\theta)$ values.

The wave-propagation model based on the Huygens' principle implemented in Prometheus performs a series of approximations with respect to the burning area (ellipse) such that a cell is classified as a \textit{Burned} cell in the Burn Grid output (0-1 Matrix) only if more than 50\% of its area has been covered by fire (belongs to the interior of the approximated ellipse) while in Cell2Fire a cell is always completely available or burned. In addition, an ignition point represents a complete cell in Cell2Fire (an area) while it is just a vertex/point for Prometheus, defining two different (but consistent) starting points for the fire spread evolution. This approximation based on discrete cells improves as the cell size decreases.

\begin{table}[h!]
\centering
\begin{adjustbox}{max width=0.7\textwidth}
\begin{tabular}{@{}c|cc|cc@{}}
                      & \multicolumn{2}{c|}{\textbf{Sub-1}}                & \multicolumn{2}{c}{\textbf{Sub-2}}                 \\ \hline
\textbf{Hour}         & \textbf{1 - MSE {[}\%{]}} & \textbf{SSIM {[}\%{]}} & \textbf{1 - MSE {[}\%{]}} & \textbf{SSIM {[}\%{]}} \\ \hline
\textbf{1}            & 99.75                     & 93.59                  & 99.98                     & 95.44                  \\
\textbf{2}            & 96.75                     & 77.40                  & 99.94                     & 95.01                  \\
\textbf{3}            & 97.75                     & 74.03                  & 95.56                     & 85.53                  \\
\textbf{4}            & 97.50                     & 78.70                  & 95.38                     & 84.41                  \\
\textbf{5}            & 96.75                     & 79.61                  & 96.50                     & 82.25                  \\
\textbf{6}            & 96.00                     & 73.96                  & 94.03                     & 75.01                  \\
\hline
\textbf{AVG {[}\%{]}} & 97.42                     & 79.55                  & 96.90                     & 86.25                  \\ 
\end{tabular}
\end{adjustbox}
\caption{Sub-Instances accuracy and structural similarity index measure values per hour (6 hours evolution)}
\label{detresults}
\end{table}

The two fire growth models are consistent and correlated in that they produce very similar fire scars for the test instances as seen in Figure \ref{sub2}.



\subsubsection{British Columbia wildfire set}
The British Columbia instances set contains five different regions --- ArrowHead (265,536 ha.), Revelstoke (391,314 ha.), Mica Creek (348,404 ha.), Glacier Natural Park (559,746 ha.), and Central Kootenay (494,665 ha.) --- of the province. For each area, two fires with random ignition points and 24-hours stream weather conditions based on the historical weather dataset from the Climate Information Section of the Agriculture and Forestry site of Alberta, Canada, and data from the Yaha Tinda Auto station (coordinates: 51.6547 , -115.3617) are compared. These instances are provided with BurnP3 \url{http://www.firegrowthmodel.ca/burnp3/software_e.php}. We generated subsets of the large forests, simulated fires in Prometheus using 24 hours weather scenarios --- using historical data from the zone --- and then we compare the fire scars from Prometheus with the ones obtained with Cell2Fire.

\begin{figure}[h!]
    \centering
    \begin{subfigure}{.5\textwidth}
      \centering
      \includegraphics[scale=0.3]{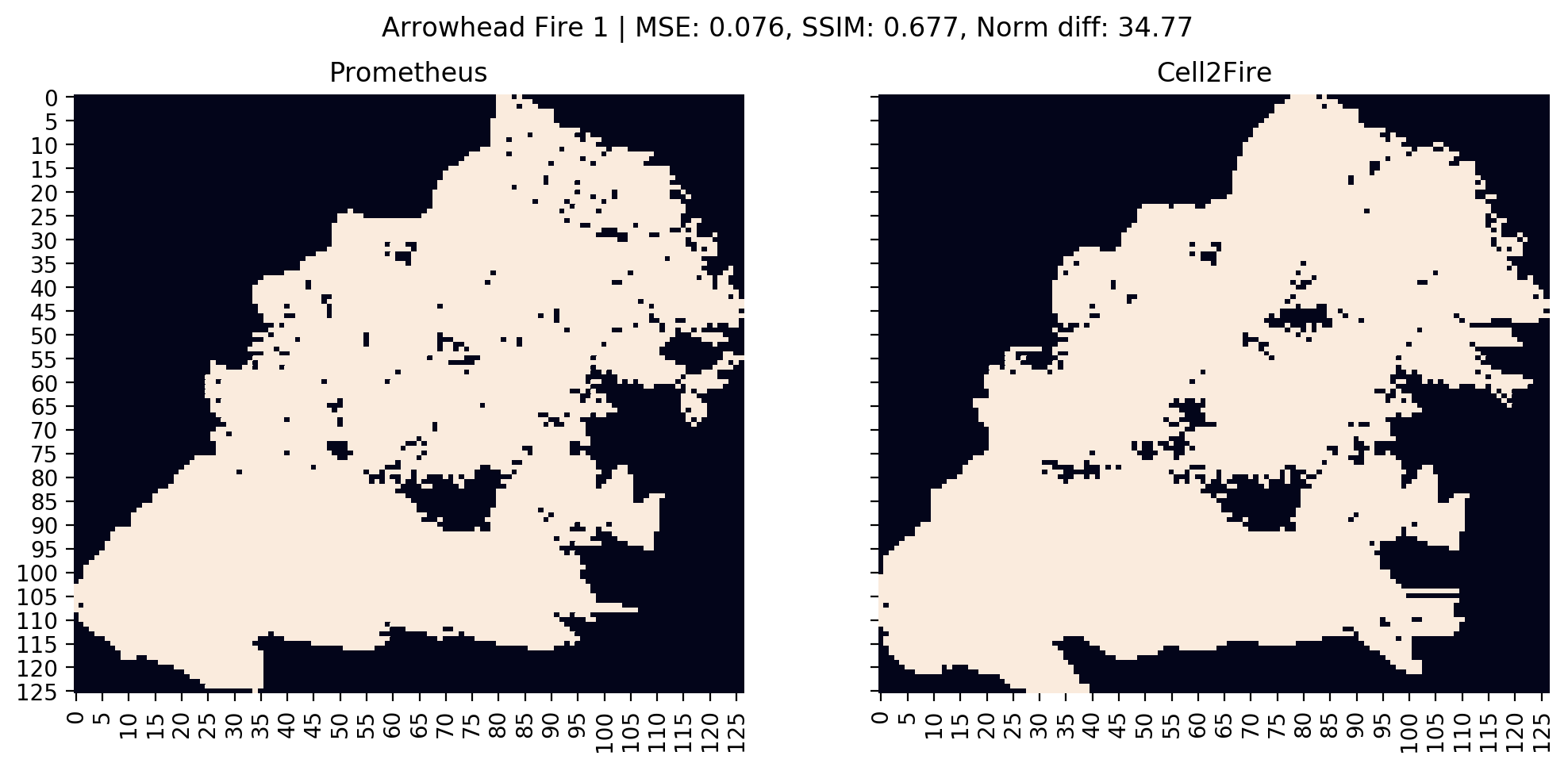}
      \label{AF1}
    \end{subfigure}%
    \begin{subfigure}{.55\textwidth}
       \centering
       \includegraphics[scale=0.3]{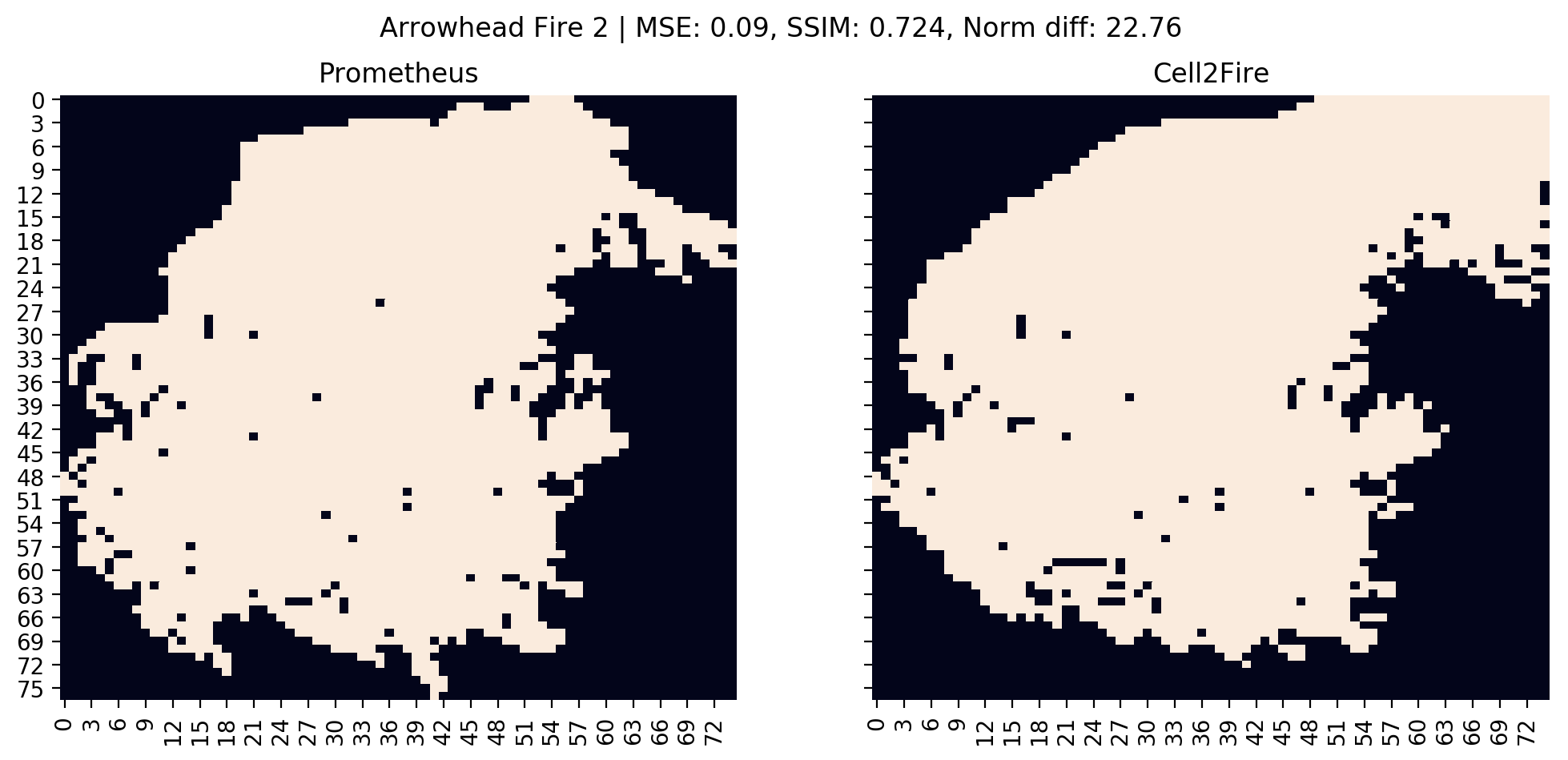}
       \label{AF2}
    \end{subfigure}
    \label{AH}
    \vspace{0.3cm}
    \begin{subfigure}{.5\textwidth}
      \centering
      \includegraphics[scale=0.3]{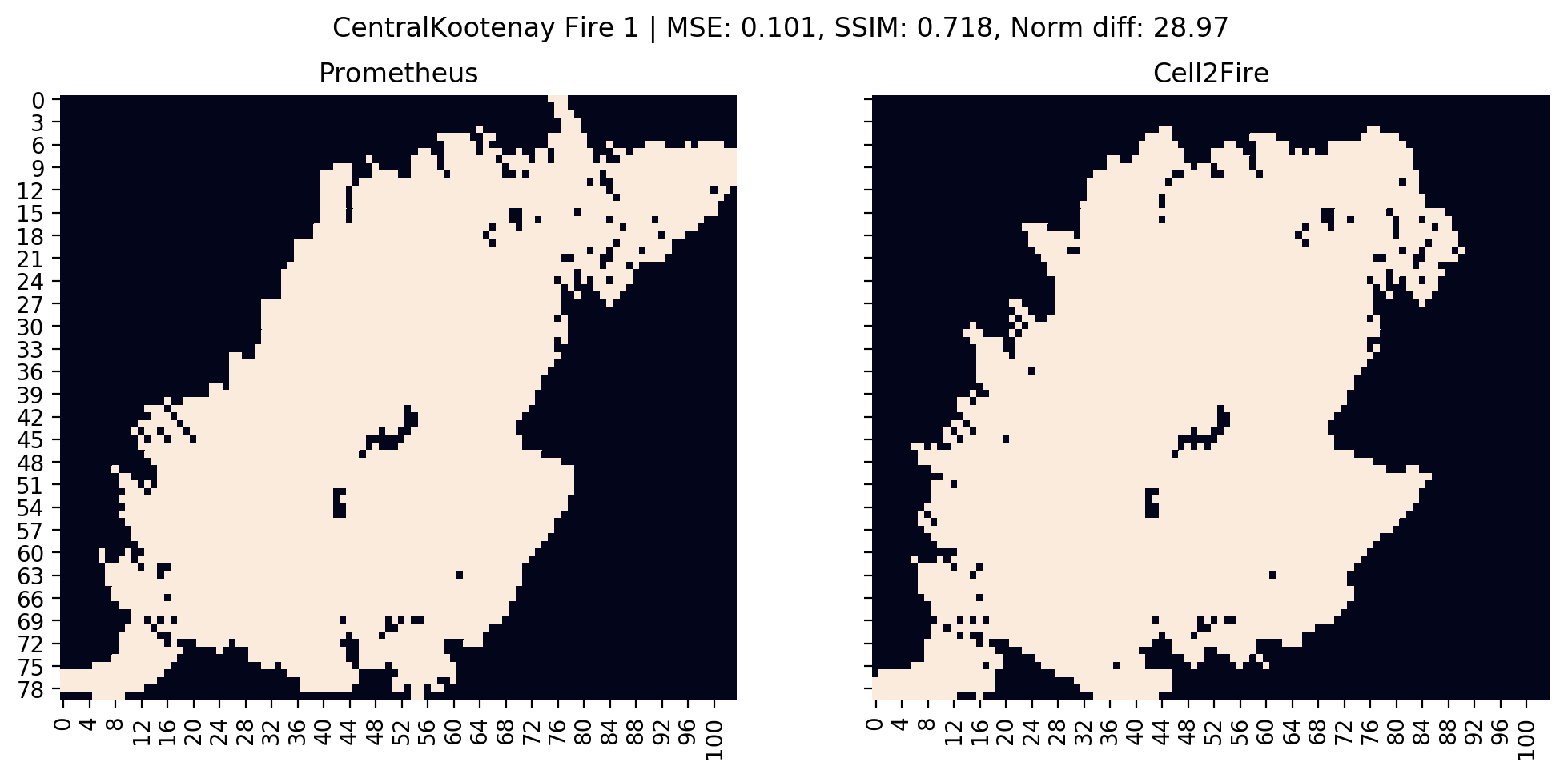}
      \label{CK1}
    \end{subfigure}%
    \begin{subfigure}{.55\textwidth}
       \centering
       \includegraphics[scale=0.3]{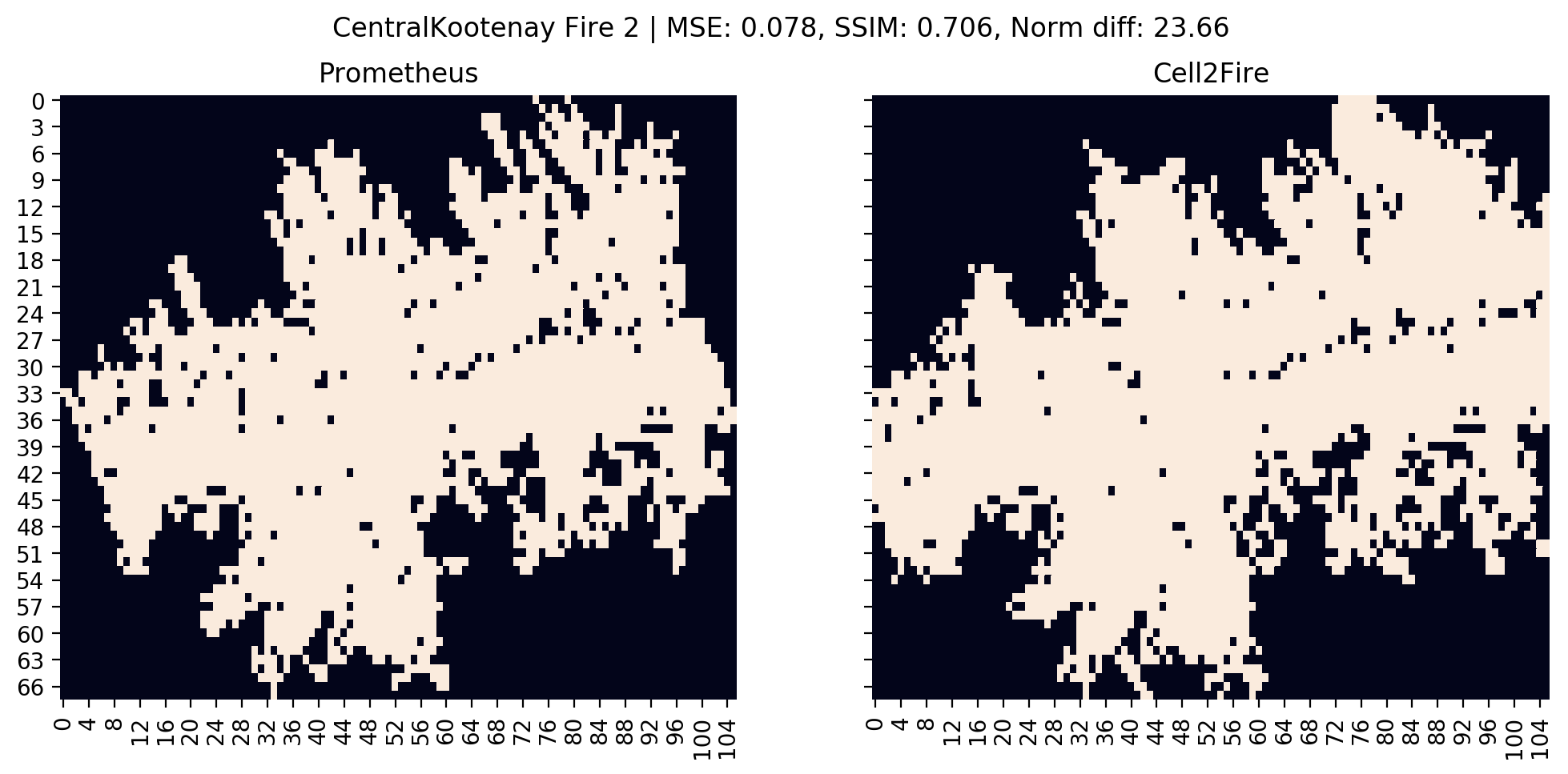}
       \label{CK2}
    \end{subfigure}
    \label{CK}
    \vspace{0.3cm}
    \begin{subfigure}{.5\textwidth}
      \centering
      \includegraphics[scale=0.3]{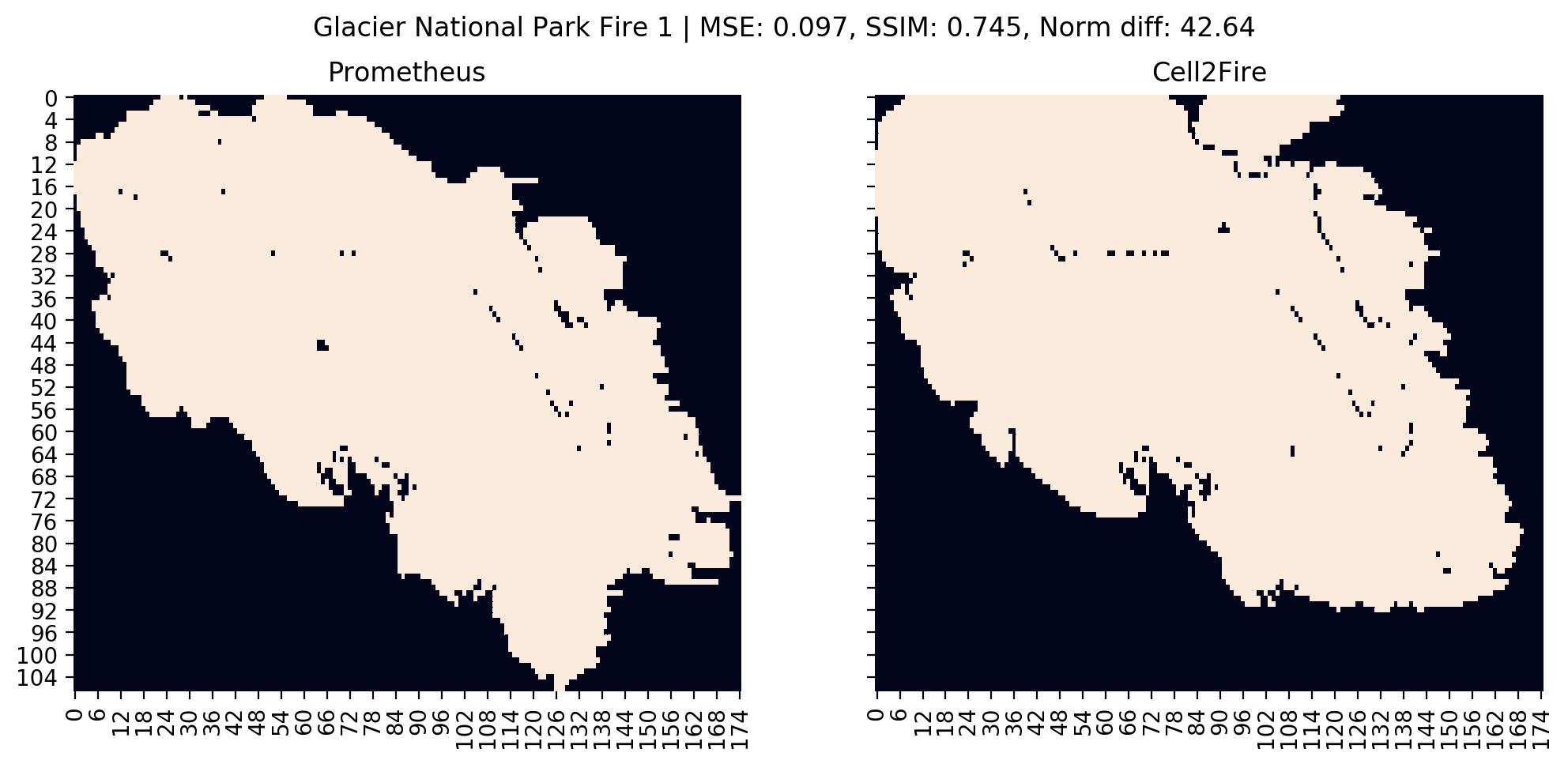}
      \label{G1}
    \end{subfigure}%
    \begin{subfigure}{.55\textwidth}
       \centering
       \includegraphics[scale=0.3]{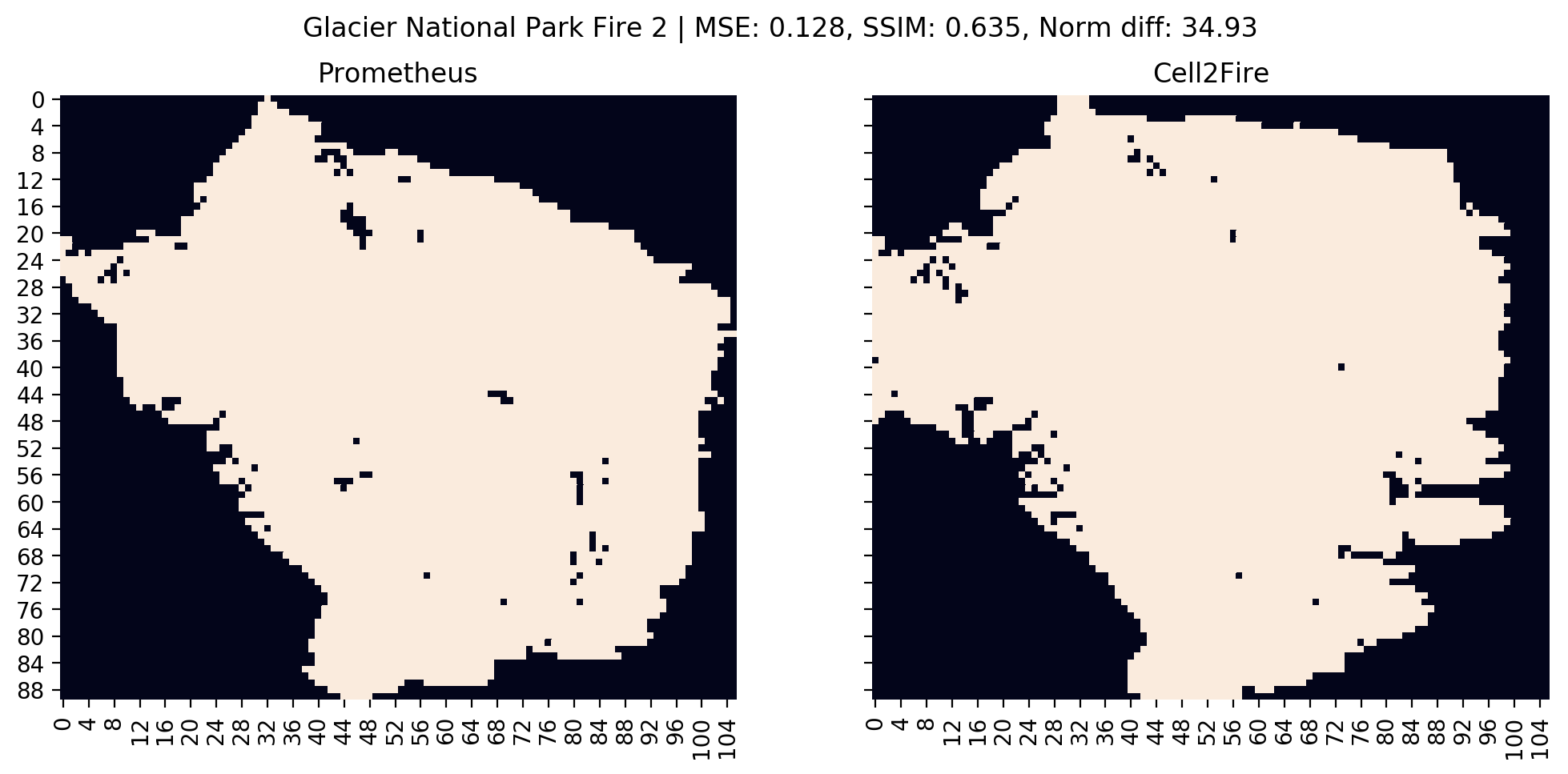}
       \label{G2}
    \end{subfigure}
    \label{G}
    \vspace{0.3cm}
    \begin{subfigure}{.5\textwidth}
      \centering
      \includegraphics[scale=0.3]{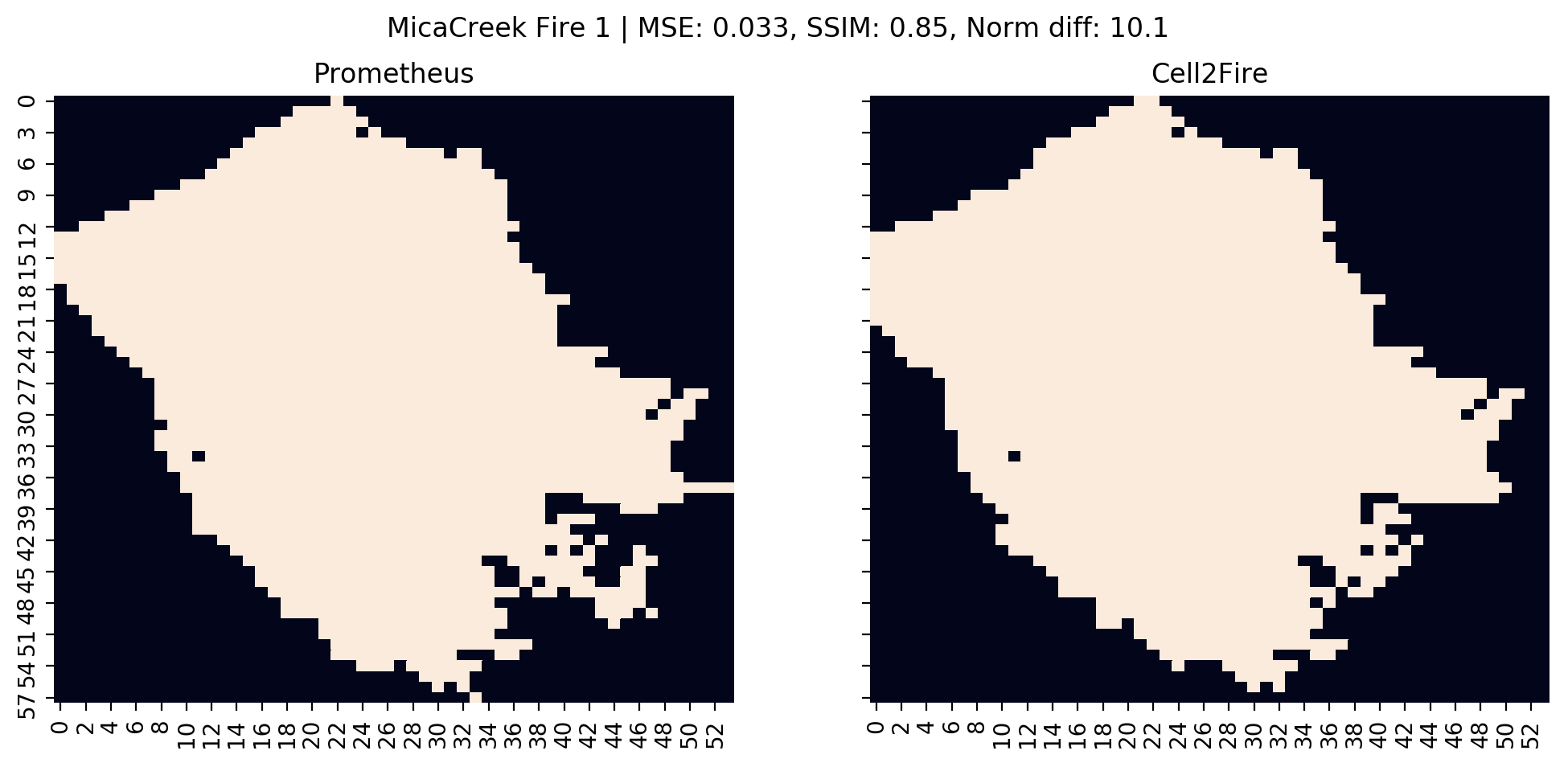}
      \label{MC1}
    \end{subfigure}%
    \begin{subfigure}{.55\textwidth}
       \centering
       \includegraphics[scale=0.3]{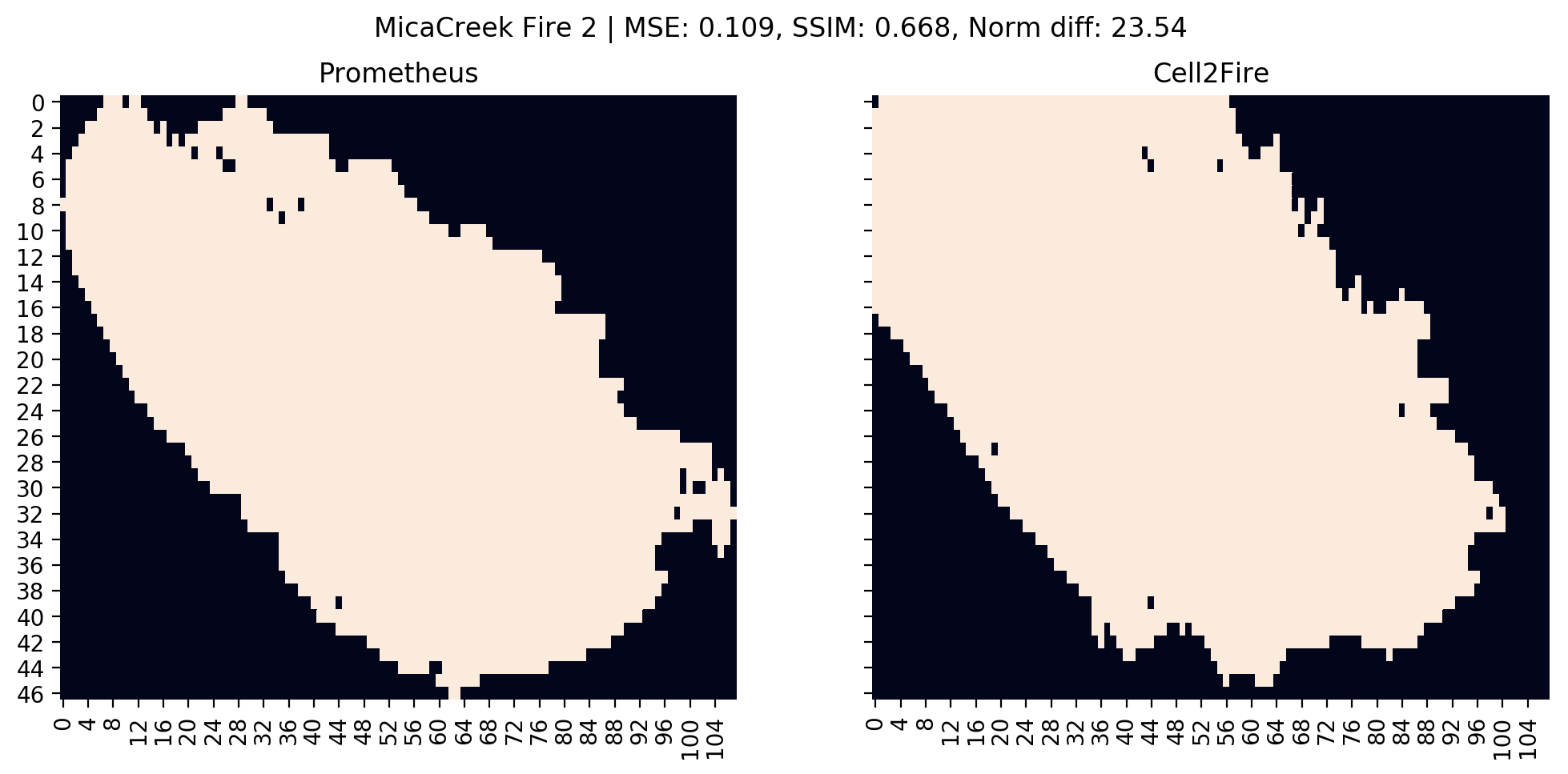}
       \label{MC2}
    \end{subfigure}
    \label{MC}
    \vspace{0.3cm}
    \begin{subfigure}{.5\textwidth}
      \centering
      \includegraphics[scale=0.3]{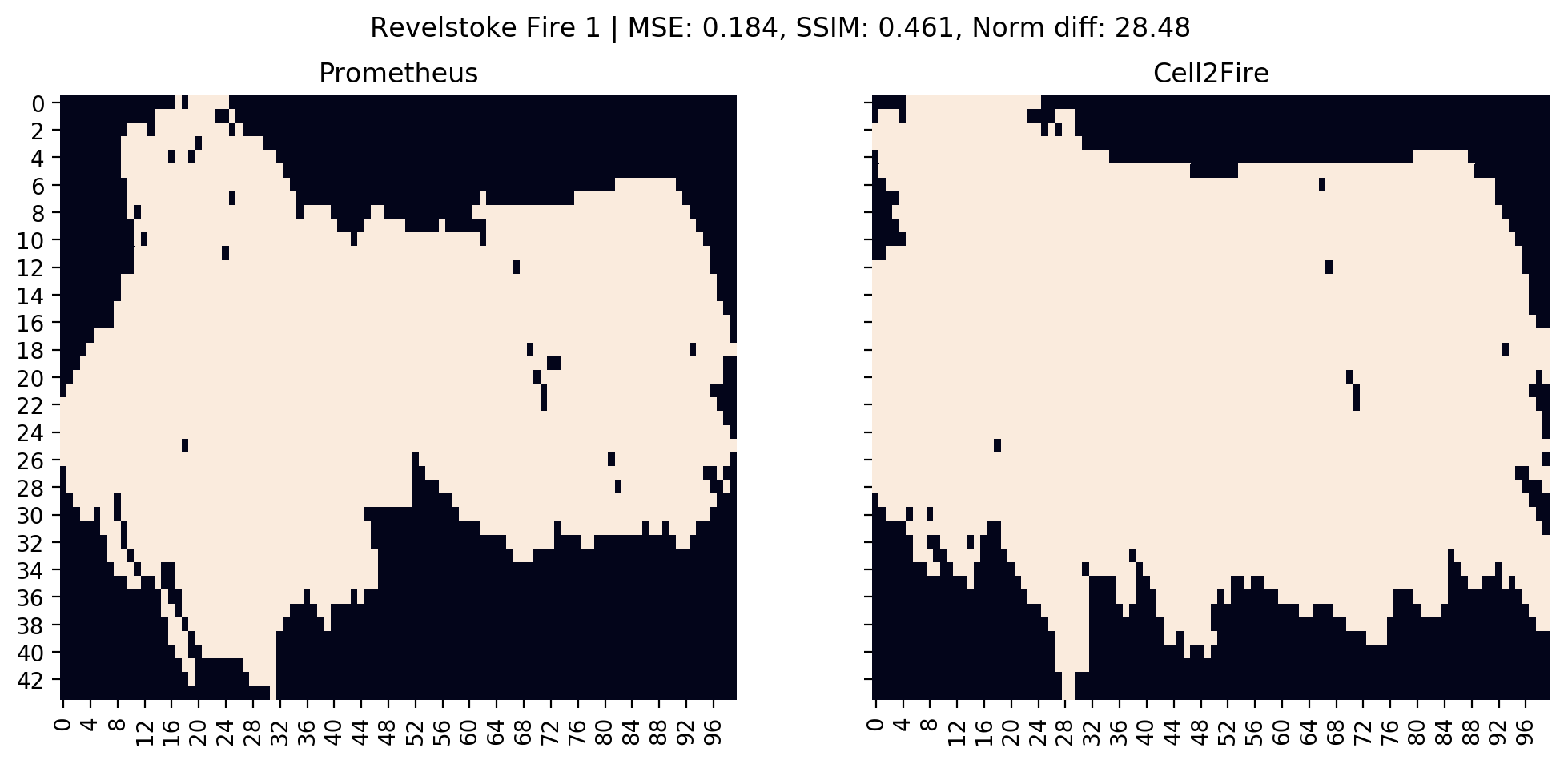}
      \label{RS1}
    \end{subfigure}%
    \begin{subfigure}{.55\textwidth}
       \centering
       \includegraphics[scale=0.3]{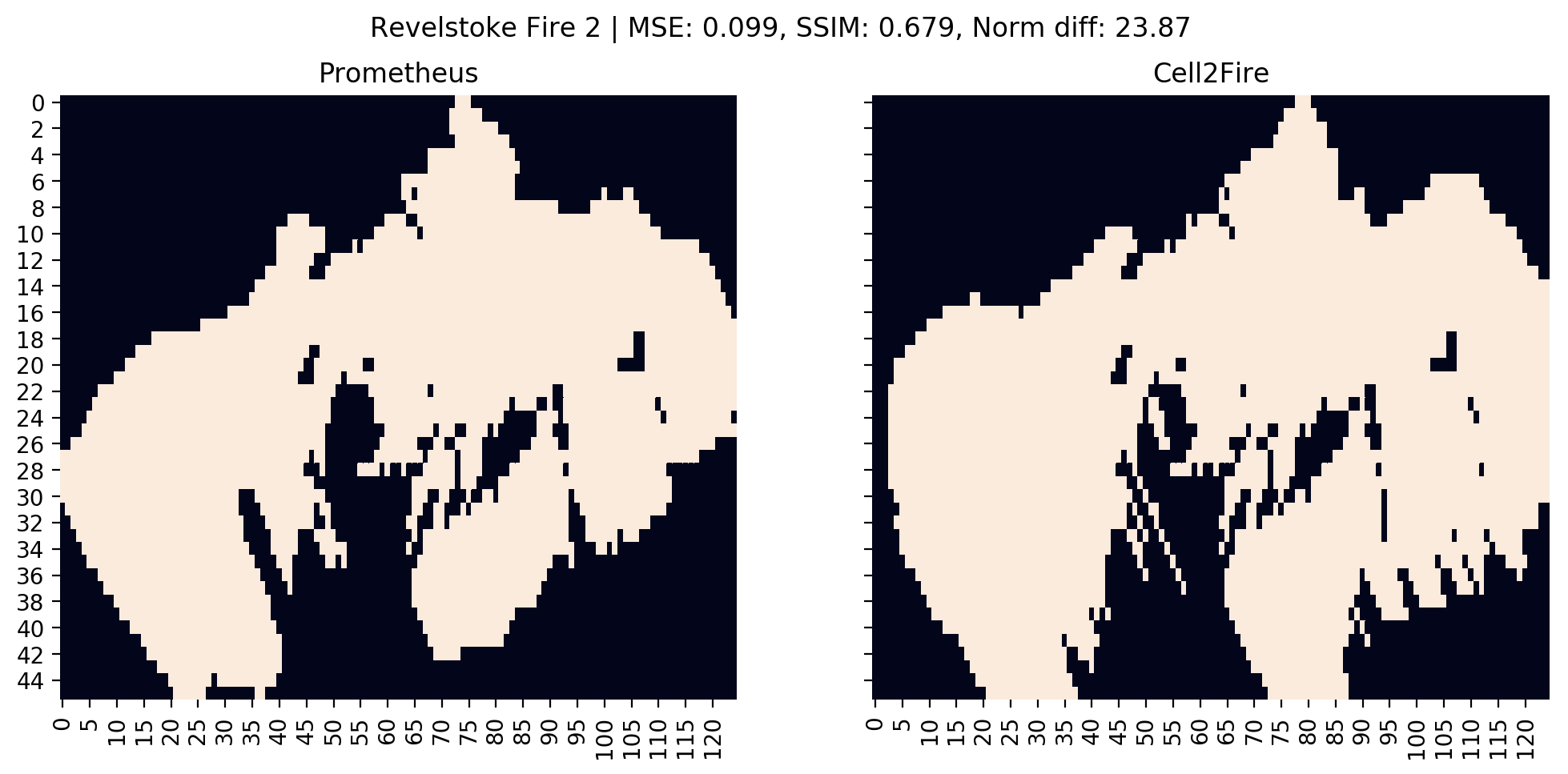}
       \label{RS2}
    \end{subfigure}
    \label{RS}
    \caption{British Columbia wildfire instances. From the final scars and statistics (left side Prometheus, right side Cell2Fire), it can be seen how accurate is Cell2Fire w.r.t. simulated scars from Prometheus, reaching $\overline{MSE} = 0.0995$, $\overline{SSIM} = 0.6863$, $\overline{\delta}_{norm} = 16.347$.}
    \label{BCOl}
    \end{figure}

    The final fire scars and performance metrics --- focusing in the affected area of the instance for easier visualization --- obtained for the 10 simulated wildfires in both Prometheus (columns 1 and 3) and Cell2Fire (columns 2 and 4) can be seen in Figure \ref{BCOl}. Results indicate the high similarity between the scars, obtaining good performance across the main three metrics (Table \ref{BCSummary}) for all forests. This validates Cell2Fire for different fuel types, landscapes, and weather scenarios, being able to approximate the results of a state-of-the-art simulator like Prometheus.
    
    Different ignition points and weather scenarios were tested on these landscapes, obtaining similar results in terms of the main performance metrics. A similar pattern was observed with respect to the hourly evolution of the scars. 
    \begin{table}[h!]
    \centering
    \begin{tabular}{@{}cccc@{}}
    \toprule
    \textbf{}                          & \textbf{$MSE$} & \textbf{$SSIM$} & \textbf{$\delta_{norm}$} \\ \midrule
    \multicolumn{1}{c|}{\textbf{Mean}} & 0.09           & 0.68            & 27.36                    \\
    \multicolumn{1}{c|}{\textbf{Std}}  & 0.04           & 0.09            & 8.88                     \\
    \multicolumn{1}{c|}{\textbf{Max}}  & 0.18           & 0.85            & 42.64                    \\
    \multicolumn{1}{c|}{\textbf{Min}}  & 0.03           & 0.46            & 10.10                    \\ \bottomrule
    \end{tabular}
    \caption{British Columbia simulations summary statistics when comparing the simulated final fire scars from Prometheus and Cell2Fire. $\delta_{norm} = || X- Y ||$ where $X$ and $Y$ are the binary BurnGrids matrices obtained from both simulators.}
    \label{BCSummary}
    \end{table}

\subsubsection{Dogrib fire instance \label{sec:dogrib}}
\begin{figure}[h!]
	\centering
    \includegraphics[scale=0.8]{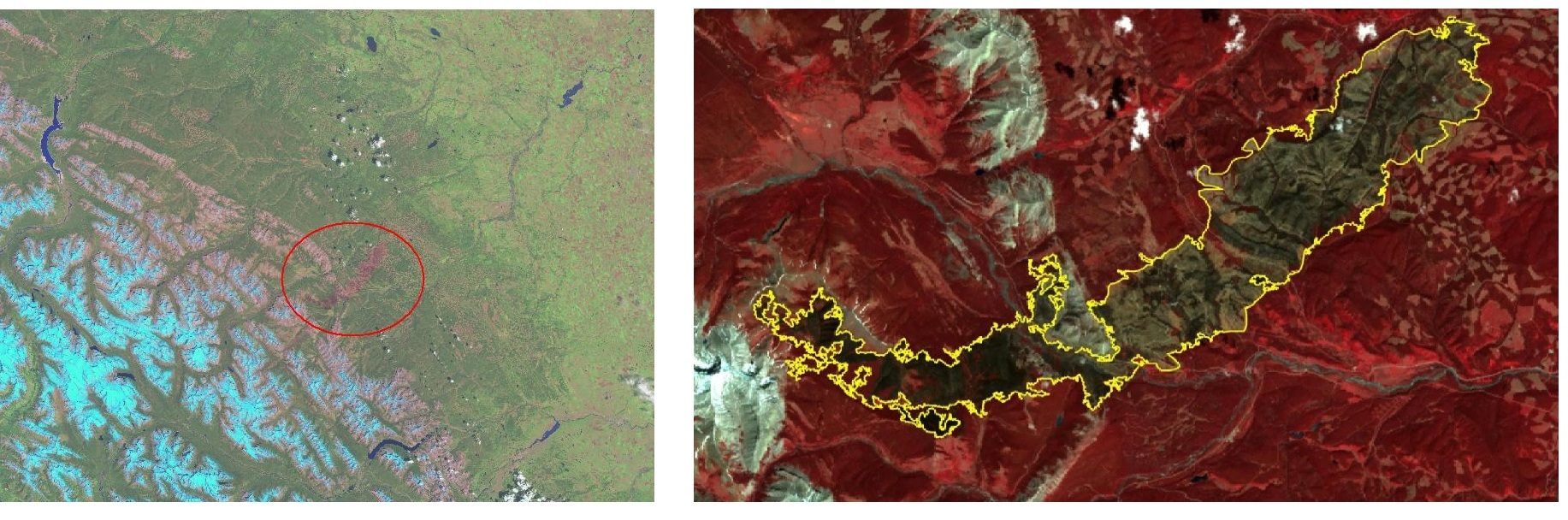}
    \caption{Dogrib fire perimeter registered on June 22, 2002 from Landsat.}
    \label{DogribLandsat}
\end{figure}

  The Dogrib fire \citep{Prometheus} started on September 25, 2001 the province of Alberta. The fire was detected late in the afternoon on September 29 and assessed early the next day at 70 ha. in size. Suppression action began early on October 1. The fire was 828 ha. and out of control on October 15. A wind event resulted in a major fire run on October 16. Local terrain funneled wind flow along the Red Deer River and through a gap in the surrounding mountains. This pushed the fire east along the river valley. The fire jumped the Red Deer River and a road and then resumed spreading in a northeast direction. The final fire size was 10,216 ha. The October 16 fire run accounted for ninety percent of the total area burned and resulted in high to very high burn severities. In Figure \ref{DogribLandsat} we can see the Dogrib fire perimeter and burn severity as detected by Landsat.

    \newpage
    \clearpage

We chose to model this particular fire due to the large amount of documentation and real data available --- weather conditions recorded from The Yaha Tinda Automatic station, and demographic/topographic data collected from the area --- as well as the fact that it contained a representative set of different fuel-types documented in the Canadian FBP system. We divided the landscape into 79,611 $100 \times 100;\ [m^2]$ cells, used the Dogrib fire's ignition point located at (51.652876\degree, -115.477908\degree) and started spreading the fire on October 16, 2001, 13:00 hrs. As discussed in the previous section, the ignition point is translated into an ignition area (cell) in Cell2Fire containing its coordinates.

\begin{figure}[h!]
	\centering
    \includegraphics[scale=0.23]{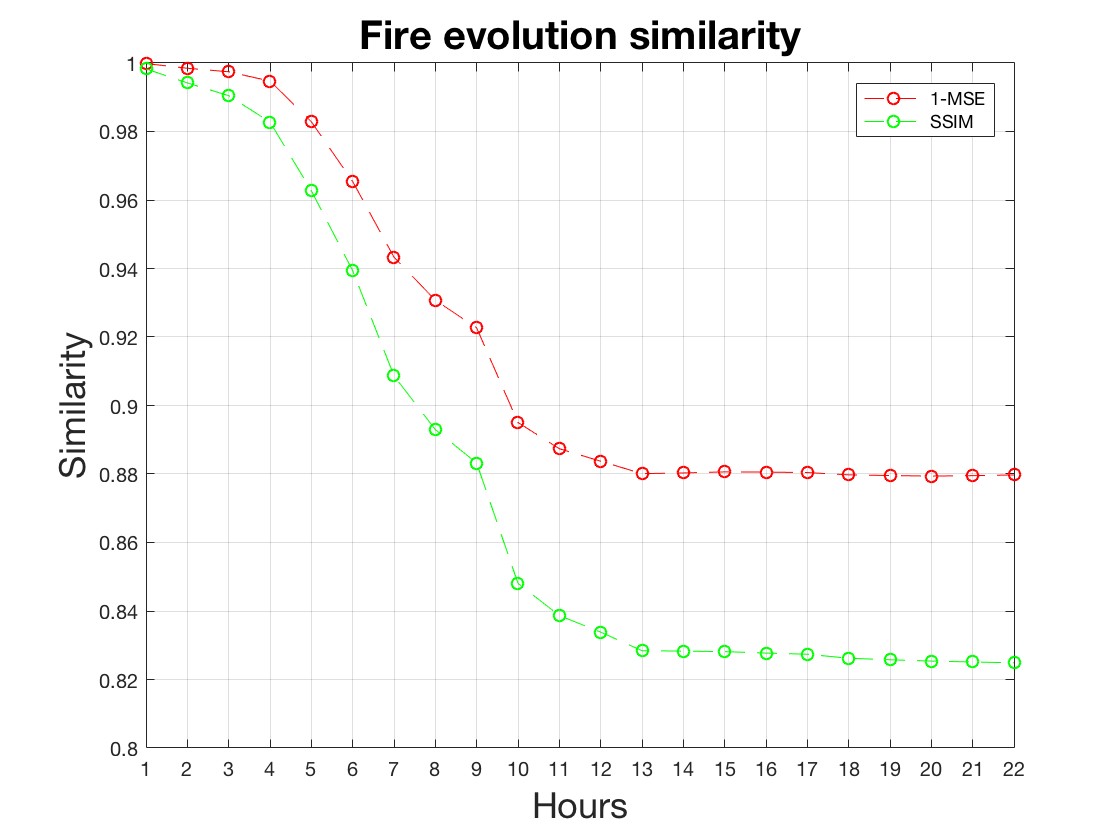}
    \caption{Dogrib MSE and SSIM hourly evolution (22 hours).}
    \label{compare_dogrib}
\end{figure}

The comparison of the performance metrics after 22 hours of fire growth is shown in Figure~\ref{compare_dogrib} . Here, it can be seen that Cell2Fire has a very similar evolution with respect to the wave-front approach, obtaining good performance when compared with Prometheus, not exceeding a 20\% difference in both measurements. An average of 87.91\% of structural similarity and a global average of 91.82\% of accuracy (1-$MSE$) are obtained during the 22 hours of active fire growth. A clear pattern can be seen in the graph where both performance metrics start very high and remain stable during the first 4 hours, then they show a significant negative slope between hours 4 to 11, and finally reach a steady state for the rest of the simulation. The explanation behind these results is clear: during the initial 4 hours of the fire, similar fire growth occurs due to weather conditions that are not extreme, resulting in diminished model differences (wave-front and cellular-automata), however, weather factors between hours 4 to 11 contain the most extreme conditions (strong wind speed, high temperature, low precipitation levels, etc.) magnifying the fire growth differences/approximations between both approaches in terms of the number of burning cells per hour (fire scar). After hour 11, differences between fire scars tend to be very stable due to the lack of new extreme weather episodes, but keeping the structural differences --- mainly in terms of the current perimeter of the fire --- in the fire scars obtained in the previous hours. 

In addition, some of the structural differences between the generated fire scars can be explained in part by the extra modeling features included in Prometheus but not in Cell2Fire such as \textit{Breaching} where non-fuel grid cells or linear fuel breaks fail to stop an advancing fire front, a feature that is not currently included in Cell2Fire.

\begin{figure}[h!]
	\centering
    \includegraphics[scale=0.90]{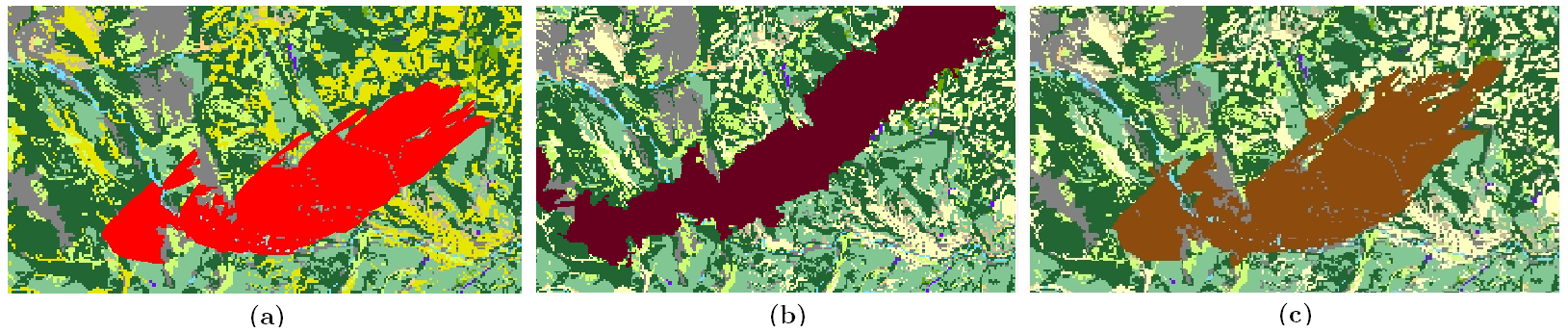}
    \caption{a) Prometheus fire scar obtained for the region of Dogrib, Canada, comparison with the b) real fire projected into grid format in 2002 and c) Cell2Fire final output.}
    \label{DogribScar}
\end{figure}

Detailed results are presented in Table \ref{DogribEvoT} where both ($1-MSE$) and $SSIM$ values can be seen for the full 22 simulated hours. We conclude that Cell2Fire produces results that are similar to those produced by Prometheus, obtaining similar final fire scars as seen in Figure \ref{DogribScar} where the simulated fires and the real satellite images are shown.  

\begin{table}[h!]
\centering
\begin{adjustbox}{max width=0.8\textwidth}
\centering 
\begin{tabular}{c|c|c}
\textbf{Hour}         & \textbf{1-MSE {[}\%{]}} & \textbf{SSIM {[}\%{]}} \\ \hline
\textbf{1}            & 99.98                   & 99.83                  \\
\textbf{2}            & 99.85                   & 99.42                  \\
\textbf{3}            & 99.74                   & 99.05                  \\
\textbf{4}            & 99.47                   & 98.28                  \\
\textbf{5}            & 98.29                   & 96.27                  \\
\textbf{6}            & 96.54                   & 93.94                  \\
\textbf{7}            & 94.33                   & 90.86                  \\
\textbf{8}            & 93.08                   & 89.30                  \\
\textbf{9}            & 92.26                   & 88.32                  \\
\textbf{10}           & 89.50                   & 84.81                  \\
\textbf{11}           & 88.74                   & 83.87                  \\ \hline
\textbf{AVG {[}\%{]}} & 95.62                   & 93.08  \\               
\end{tabular}
\quad
\begin{tabular}{c|c|c}
\textbf{Hour}         & \textbf{1-MSE {[}\%{]}} & \textbf{SSIM {[}\%{]}} \\ \hline
\textbf{12}           & 88.37                   & 83.38                  \\
\textbf{13}           & 88.01                   & 82.84                  \\
\textbf{14}           & 88.03                   & 82.83                  \\
\textbf{15}           & 88.06                   & 82.81                  \\
\textbf{16}           & 88.06                   & 82.77                  \\
\textbf{17}           & 88.04                   & 82.74                  \\
\textbf{18}           & 87.98                   & 82.62                  \\
\textbf{19}           & 87.96                   & 82.58                  \\
\textbf{20}           & 87.94                   & 82.54                  \\
\textbf{21}           & 87.95                   & 82.51                  \\
\textbf{22}           & 87.97                   & 82.49                  \\ \hline
\textbf{AVG {[}\%{]}} & 88.03                   & 82.74 \\

\end{tabular}
\end{adjustbox}
\caption{Dogrib accuracy and structural similarity index measure values per hour (22 hours evolution).}
\label{DogribEvoT}
\end{table}

\section{Conclusions}

Cell2Fire provides numerous opportunities for researchers interested
in incorporating fire growth in their models for strategic harvest planning and fuel
management. We are currently using it in ongoing projects.
Because it is open-source and modular it
lends itself to customization as needed.  The simulator
is fast and scales well in parallel computing environments
so it is well-suited for use with large forests and large studies
that may require many simulation runs.

Using the FBP fire spread model, we have compared the simulated fire scars with those produced by the state-of-the-art simulator, Prometheus. Other fire spread
models can be employed instead, which extends the range of
environments where Cell2Fire can be used and also allows
for comparison of fire spread models when used in a growth
simulator. In addition to supporting stochastic ignition
and weather, the simulator also supports modeling of the
uncertainty in the rate of spread.

By adding a highly parallelizable, open-source fire growth simulator to the
tool set available, we hope to provide transparent support for ongoing
research.


\section*{Acknowledgements}
We thank B.M. Wotton and D. Boychuck for their helpful comments and suggestions. This research has been supported by Complex Engineering Systems Institute (CONICYT-PIA-FB0816) and FONDECYT 1191531.

\bibliography{Bibliography.bib}

\begin{figure}[h!]
	\centering
    \includegraphics[scale=0.52]{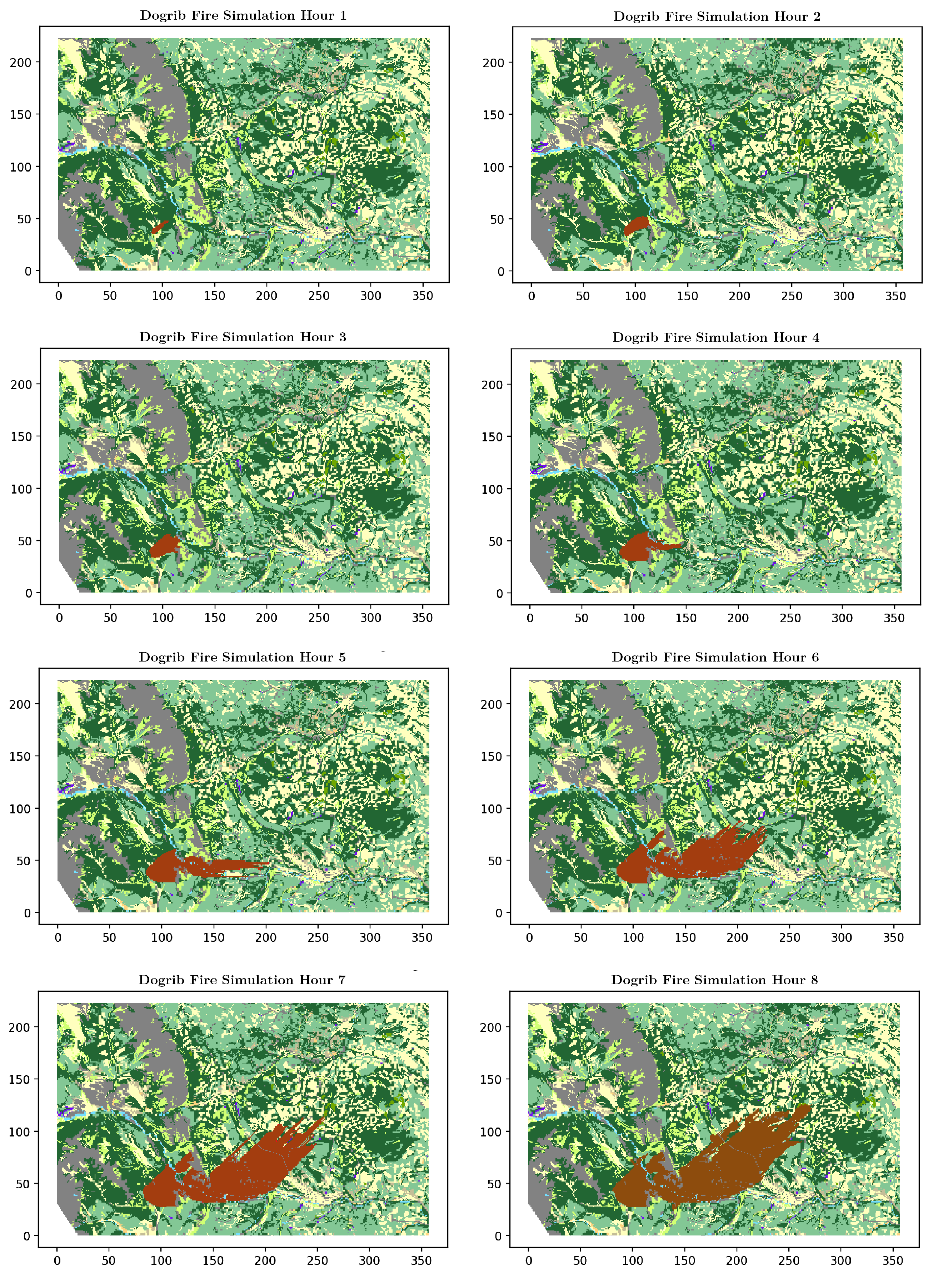}
    \caption{Dogrib fire scar evolution for the first 8 hours of the simulation inside Cell2Fire.}
    \label{DogribScarEvolution}
\end{figure}

\newpage
\appendix
\section{Parallel performance analysis}
\subsection{Methodology}
The high complexity of the simulation scheme poses a challenge for measuring the performance of our implementation: each fire has a large number of parameters and specific characteristics that lead to different outcomes and thus, the potential for parallelizing. Thus, different fuel types, weather streams, forest structures, and/or ignition points could lead to completely different fire dynamics (e.g., number of simultaneous burning cells, number of burned hectares), and hence, to different serial/parallel performance. In order to account for this, multiple instances/forests based on real fuel, weather and topographic data are generated, averaging their results, and comparing the performance of Cell2Fire.

Performance is measured by calculating both the strong and weak scaling efficiencies --- as well as speedup factors --- obtained for different experimental instances ranging from sizes (number of cells inside the forest) $n \in [4-1M]$. Summary plots are generated in order to visualize the performance of our parallel implementation.

\subsection{Instances: Data}
Two weather files from weather stations located in Canada containing all relevant inputs for 7 and 36 hours are used for all instances. Each set of experiments for $n$ cells uses the same ignition points for comparison purposes, starting the fire dynamic at the same hour for 1-minute precision clock step-size. Larger instances are generated by mixing locations using real data gathered from Canadian forests. In addition, homogeneous instances (same fuel type for all cells) are included in each experimental set for comparison purposes.

\subsection{Hardware \& Software}
The optimization and parallelization of Cell2Fire are developed for a specific hardware and run-time environment for the National Energy Research Scientific Computing Center (NERSC). In addition, tests have been performed in a common daily-use laptop for comparison purposes. All experiments, benchmarks, and performance results are implemented using the following hardware and software:

\textbf{1. NERSC's Cori supercomputer: Phase I} 
\begin{itemize}
	\item  Intel® Xeon™ Processor E5-2698 v3 ("Haswell") at 2.3 GHz (32 cores per node)
    \item 64 KB 8-way L1, 256 KB 8-way set L2, and 40 MB 20-way set L3 cache (shared per socket) 
    \item SUSE Linux version 4.4.74-92.38-default. Built with g++ version 4.8.5
\end{itemize}
\textbf{2. Daily-use laptop} 
\begin{itemize}
	\item  Intel® Core I7 4510U at 2.0 GHz (2 cores)
	\item 64 KB 8-way set L1, 2 x 256 KB 8-way set L2, and 4 MB 16-way set L3 cache
    \item Ubuntu 16.04.2 LTS / Windows 10
\end{itemize}

\subsection{Parallel Structure}
Our algorithm contained 3 sections at each time step: (1) checking for new lightning ignitions (igniting), (2) updating the fire progress of already-burned cells and analyzing newly burned ones (sending messages), and (3) marking newly burned cells as burning (receiving messages). 

The ignition stage is very quick (less than 1\% of total execution time), with most simulations only igniting a single time at the first time step of the simulation. The sending messages stage updates the ellipse associated with every burning cell. Because we can have a large number of cells burning at once, and there are no direct dependencies on neighboring cells, this part is easily parallelizable. Each cell, in addition to updating itself, can also ``send a burning message'' to an adjacent cell. In the receiving messages stage, we analyze the ``burn messages'' sent to non-burning cells and mark them as burned if the conditions are met. This part is also potentially parallelizable, but because the number of newly burned cells at a single time-step is dwarfed by the number of currently burning cells, we found that a speedup here is of lower priority ($\approx 10\%$ of total execution time).

\subsection{Parallelization: OpenMP work-sharing}
Due to the easily parallelizable structure of our code, the most suitable approach for parallelizing its execution consists of a shared-memory approach using the well-known OpenMP API. This is an advantage since the code will be also optimized for its execution in normal desktop/laptop computers, without needing a multi-node architecture to exploit parallelism. We found that we could easily make the loop embarrassingly parallel if instead of adding to a single data structure, we add to a data structure local to the worker thread. Since we would iterate over the initial data structure to compute statistics, we found the additional complexity of ``distributing'' the data structure does not scale with grid size. In addition, different loop scheduling options were tested: dynamic, guided, auto, run-time, and static, as well as the chunk-sized block process. Following a brute force optimization approach, we were able to obtain an average of 15\%-20\% extra performance for the parallel region (see the Appendix for more details).

One final improvement we made to our parallelism was analyzing the false-sharing effect. Because we had a $vector<DS>$ to store our ``distributed'' data structures, where DS is the data structure of choice, we found there to be a bottleneck on the parallelism exploited in the problem. Upon further analysis, we found this to be false-sharing in the array of DS backing the vector. After adding padding between elements of the array --- where optimal values were obtained following a binary search optimization approach --- we achieved a significant additional speedup from our initial attempt.

\subsection{Parallel section: \% of total code}
In order to identify the potential benefits of a parallelization, we performed a detailed analysis of the execution times, breaking it into: (1) sending time (``parallel'' region), (2) receiving time, (3) ignition time, and (4) copying time. In Tables \ref{PerPar} (a), (b) we can see that the average --- across all instances --- time spent in the parallelizable region represents a 79\% of the total execution time. The other $\approx 20\%$ is divided evenly between (2) and (4), while (3) is almost negligible. These results give us a sense of the potential impact of an efficient parallel implementation in our code.

\begin{table}[!htb]
    \begin{minipage}{.5\linewidth}
      \centering
        \resizebox{0.9\columnwidth}{!}{%
      \begin{tabular}{c|ccc}
      \hline
      \multirow{2}{*}{\textbf{Instance (n)}} & \multicolumn{3}{c}{\textbf{AVG \% Time in Parallel Zone: Sending}} \\ \cline{2-4} 
                                             & \textbf{Bottom 10\%}  & \textbf{Middle 80\%}  & \textbf{Top 10\%}  \\ \hline
      \textbf{4}                             & 80.44\%               & 86.00\%               & 94.10\%            \\
      \textbf{9}                             & 74.15\%               & 82.63\%               & 93.11\%            \\
      \textbf{400}                           & 56.75\%               & 72.09\%               & 95.84\%            \\
      \textbf{1600}                          & 64.32\%               & 71.57\%               & 90.45\%            \\
      \textbf{10K}                           & 54.00\%               & 77.46\%               & 88.34\%            \\
      \textbf{50K}                           & 62.00\%               & 71.73\%               & 87.43\%            \\
      \textbf{100K}                          & 64.50\%               & 79.13\%               & 91.23\%            \\ \hline
\textbf{AVG}                           & 65.17\%               & 77.23\%               & 91.50\%            \\ \hline
        \end{tabular}}
    \end{minipage}%
    \begin{minipage}{.5\linewidth}
      \centering
      \resizebox{0.9\columnwidth}{!}{%
\begin{tabular}{c|ccc}
\hline
\multirow{2}{*}{\textbf{Instance (n)}} & \multicolumn{3}{c}{\textbf{AVG \% Time in Parallel Zone: Sending}} \\ \cline{2-4} 
                                       & \textbf{Bottom 10\%}  & \textbf{Middle 80\%}  & \textbf{Top 10\%}  \\ \hline
            \textbf{160K}                          & 72.43\%               & 87.60\%               & 88.32\%            \\
\textbf{250K}                          & 70.82\%               & 78.28\%               & 90.05\%            \\
\textbf{500K}                          & 69.94\%               & 73.70\%               & 80.19\%            \\
\textbf{1M}                            & 68.64\%               & 78.01\%               & 89.00\%            \\
\textbf{10M}                           & 60.25\%               & 80.00\%               & 92.43\%            \\
\textbf{50M}                          & 58.30\%               & 88.77\%               & 93.11\%            \\ 
\textbf{100M}                          & 52.00\%               & 90.03\%               & 95.24\%            \\ \hline
\textbf{AVG}                           & 64.63\%               & 82.34\%               & 89.76\%            \\ \hline
        \end{tabular}}
    \end{minipage} 
    \caption{Average percentage of the running time parallelizable for different instances. Upper and lower tails are included for completeness. Results obtained by simulating 20 different forests (fuel types, spatial distribution, etc.) for each size $n$ with the same weather conditions.}
    \label{PerPar}
\end{table}

It is important is to note the fact that some instances (lower tail) will experience a poor parallel performance. The reason behind this behavior is clear: certain combinations of fuel types and forests distributions lead to a significantly smaller set of simultaneous burning cells per simulation, and thus, the parallelization of (1) will not impact the overall execution times as much as we desire. Knowing this limitation, we proceed to analyze the solving times and accuracy of the simulations. 

\subsection{Running times \& speedup analysis}
\begin{figure}[h!]
    \centering
    \begin{subfigure}{.50\textwidth}
      \centering
      \includegraphics[scale=0.23]{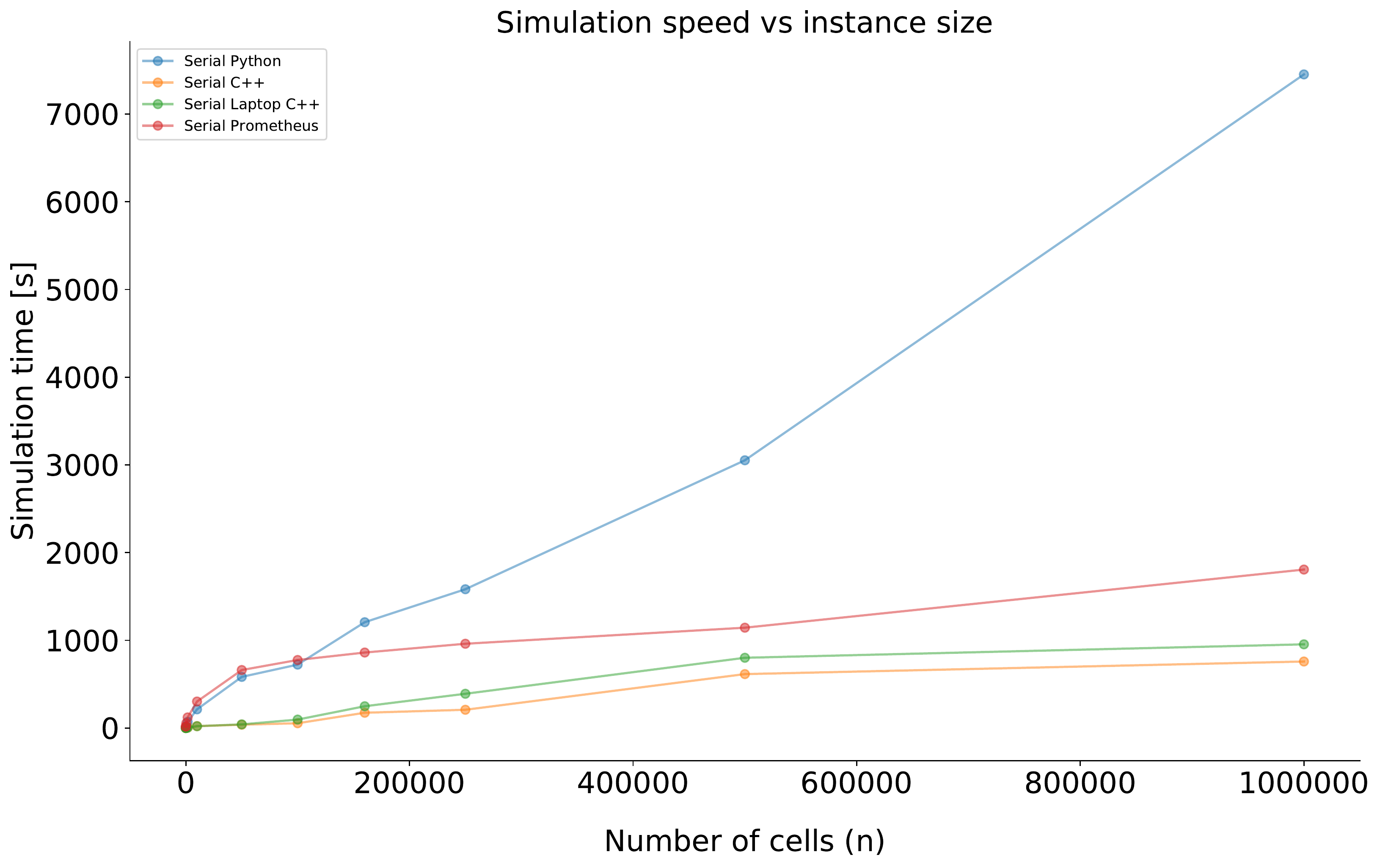}
      \caption{Running times for small instances.}
      \label{SerialAll1}
    \end{subfigure}%
    \begin{subfigure}{.50\textwidth}
       \centering
       \includegraphics[scale=0.23]{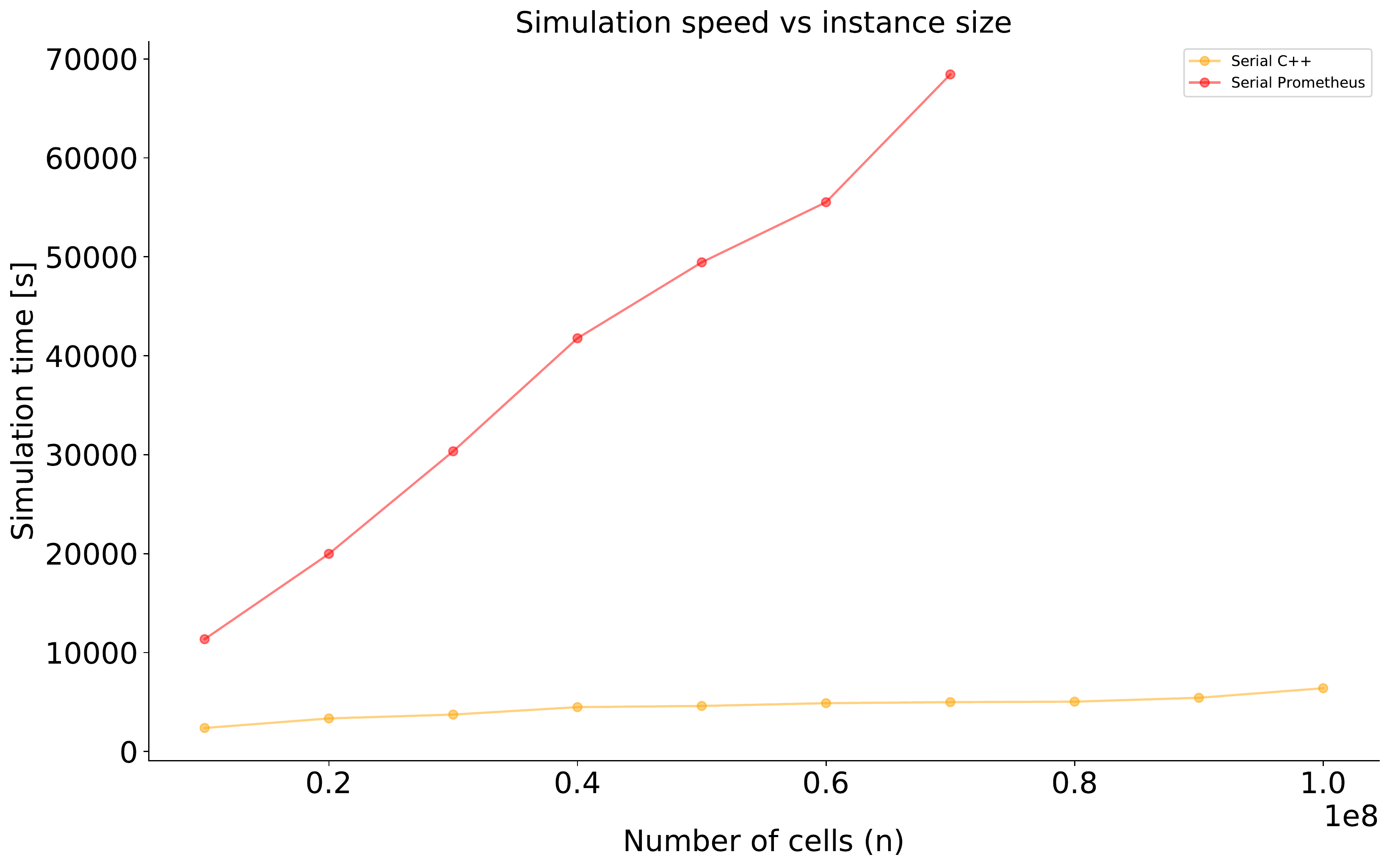}
       \caption{Running times for large instances}
       \label{SerialAll2}
    \end{subfigure}
    \caption{Running times for serial versions. Python's large instances results are omitted for visualization purposes.}
    \label{SerialAll}
    \end{figure}

Comparing the running times of our pure Python prototype and C++ implementations with the results obtained using Prometheus, we can see in Figure \ref{SerialAll} how the optimized version clearly outperforms the pure Python prototype, reaching up to 15-20x speedups when dealing with large instances. Furthermore, Cell2Fire (C++) obtains significant shorter times than Prometheus (up to 30x speedups). This is very important since wave-based simulators are performing a series of approximations when generating the final fire scar that simplifies the calculations performed under our cell-based approach, indicating that our implementation is efficient. 

It is interesting  to note that Prometheus is not able to solve the three largest instances (80M, 90M, and 100M, due to an "out of memory" error). It is therefore not suitable for massive instances. Looking at Table \ref{OpenMP1} we can see that the detailed and average speedups obtained for the small instances with the optimized OpenMP version are very good, obtaining high-performance with a certain number of threads. Performance is even better when dealing with the large instances, improving each average speedup up to an average of 16.48x when running 32 parallel threads.

\begin{table}[h!]
\centering
\caption{Speedup factors for small instances for different numbers of threads.}
\label{OpenMP1}
\resizebox{0.85\columnwidth}{!}{%
\begin{tabular}{@{}ccccccc@{}}
\toprule
\multicolumn{1}{c|}{\textbf{Instance ($n$)}} & \textbf{2 threads} & \textbf{4 threads} & \textbf{8 threads} & \textbf{16 threads} & \textbf{24 threads} & \textbf{32 threads} \\ \midrule
\multicolumn{1}{c|}{\textbf{4}}              & 0.50               & 0.40               & 1.00               & 1.00                & 1.50                & 6.00                \\
\multicolumn{1}{c|}{\textbf{9}}              & 0.71               & 2.50               & 5.00               & 3.33                & 5.00                & 5.00                \\
\multicolumn{1}{c|}{\textbf{400}}            & 3.01               & 3.26               & 3.69               & 3.66                & 3.66                & 3.69                \\
\multicolumn{1}{c|}{\textbf{1600}}           & 3.07               & 3.50               & 4.18               & 4.34                & 4.34                & 4.34                \\
\multicolumn{1}{c|}{\textbf{10000}}          & 2.24               & 2.81               & 3.64               & 4.84                & 6.12                & 8.70                \\
\multicolumn{1}{c|}{\textbf{50000}}          & 1.93               & 3.21               & 3.51               & 3.86                & 4.23                & 8.66                \\
\multicolumn{1}{c|}{\textbf{100000}}         & 1.83               & 3.72               & 4.26               & 4.35                & 5.71                & 8.38                \\
\multicolumn{1}{c|}{\textbf{160000}}         & 1.86               & 3.05               & 3.29               & 3.63                & 3.58                & 10.50               \\
\multicolumn{1}{c|}{\textbf{250000}}         & 1.85               & 3.21               & 4.76               & 5.52                & 9.12                & 11.98               \\
\multicolumn{1}{c|}{\textbf{500000}}         & 1.98               & 3.71               & 6.44               & 8.88                & 13.06               & 17.45               \\
\multicolumn{1}{c|}{\textbf{1000000}}        & 2.10               & 3.67               & 6.09               & 8.99                & 12.16               & 17.13               \\ \midrule
\textbf{AVG Speedup (OPT)}                   & 1.92               & 3.00               & 4.17               & 4.76                & 6.23                & 9.26                \\ \bottomrule
\end{tabular}}
\end{table}

As expected, even better speedups are obtained when the dealing with homogeneous forests as can be seen in the Table \ref{OpenMP2} where a summary for the large instances speedup averages is shown. From this, we can see a near optimal average speedup up to 16 threads, while reaching a great $\approx 20x$ with 32 threads.

\begin{table}[h!]
\centering
\caption{Average speedups for large instances: heterogeneous and homogeneous forests}
\label{OpenMP2}
\resizebox{0.99\columnwidth}{!}{%
\begin{tabular}{@{}lcccccc@{}}
\toprule
\textbf{}                                      & \textbf{2 threads} & \textbf{4 threads} & \textbf{8 threads} & \textbf{16 threads} & \textbf{24 threads} & \textbf{32 threads} \\ \midrule
\textbf{AVG Speedup Large Homogeneous (OPT)}   & 1.99               & 3.76               & 7.01               & 12.33               & 15.34               & 19.78               \\
\textbf{AVG Speedup Large Heterogeneous (OPT)} & 1.84               & 2.44               & 4.22               & 6.89                & 11.62               & 16.48               \\ \bottomrule
\end{tabular}}
\end{table}

\subsection{Strong Scaling}
\begin{figure}[h!]
\centering
\begin{subfigure}{.50\textwidth}
  \centering
  \includegraphics[scale=0.23]{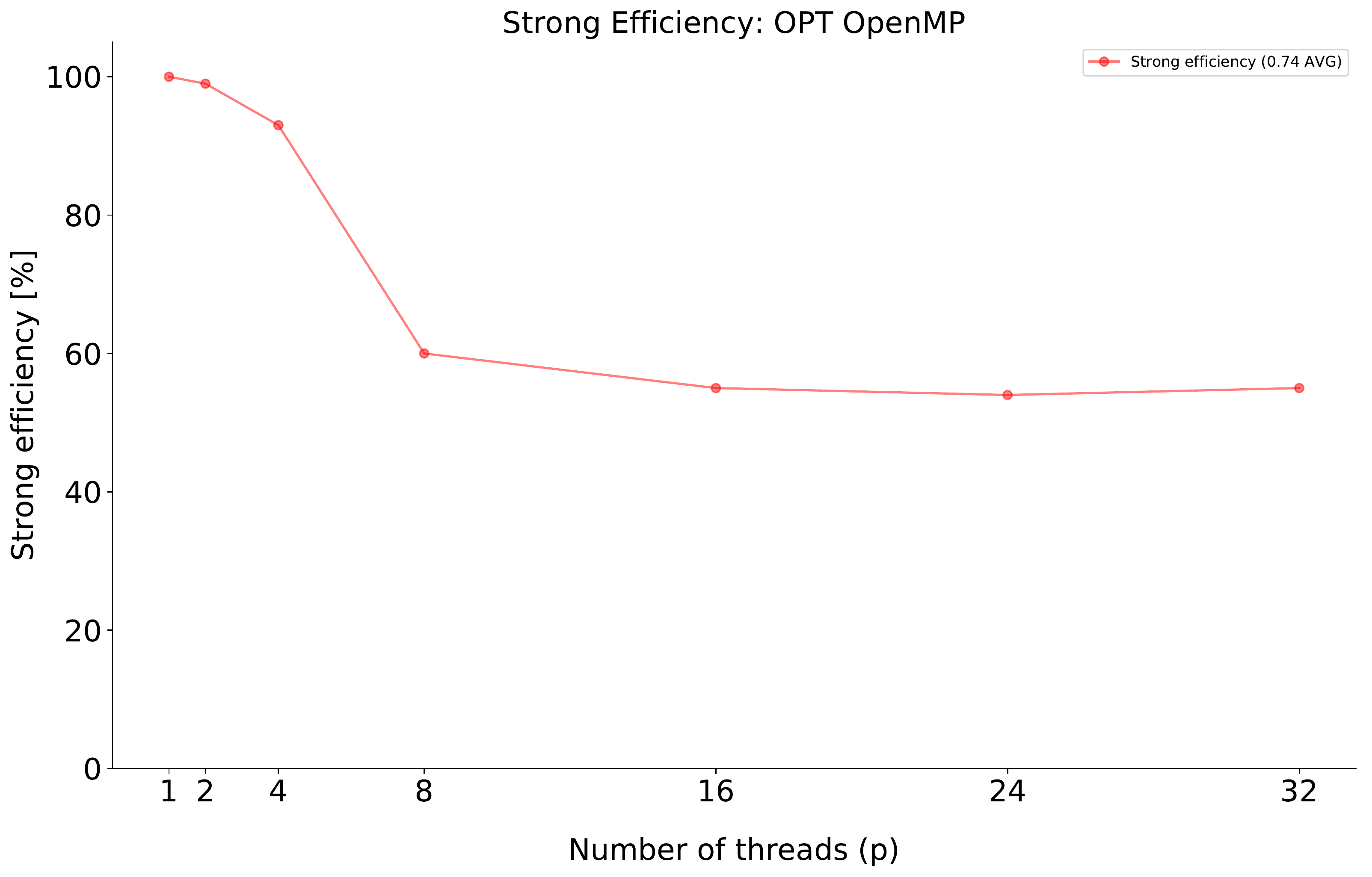}
  \caption{Strong-scaling efficiency.}
  \label{Strong1}
\end{subfigure}%
\begin{subfigure}{.50\textwidth}
   \centering
   \includegraphics[scale=0.23]{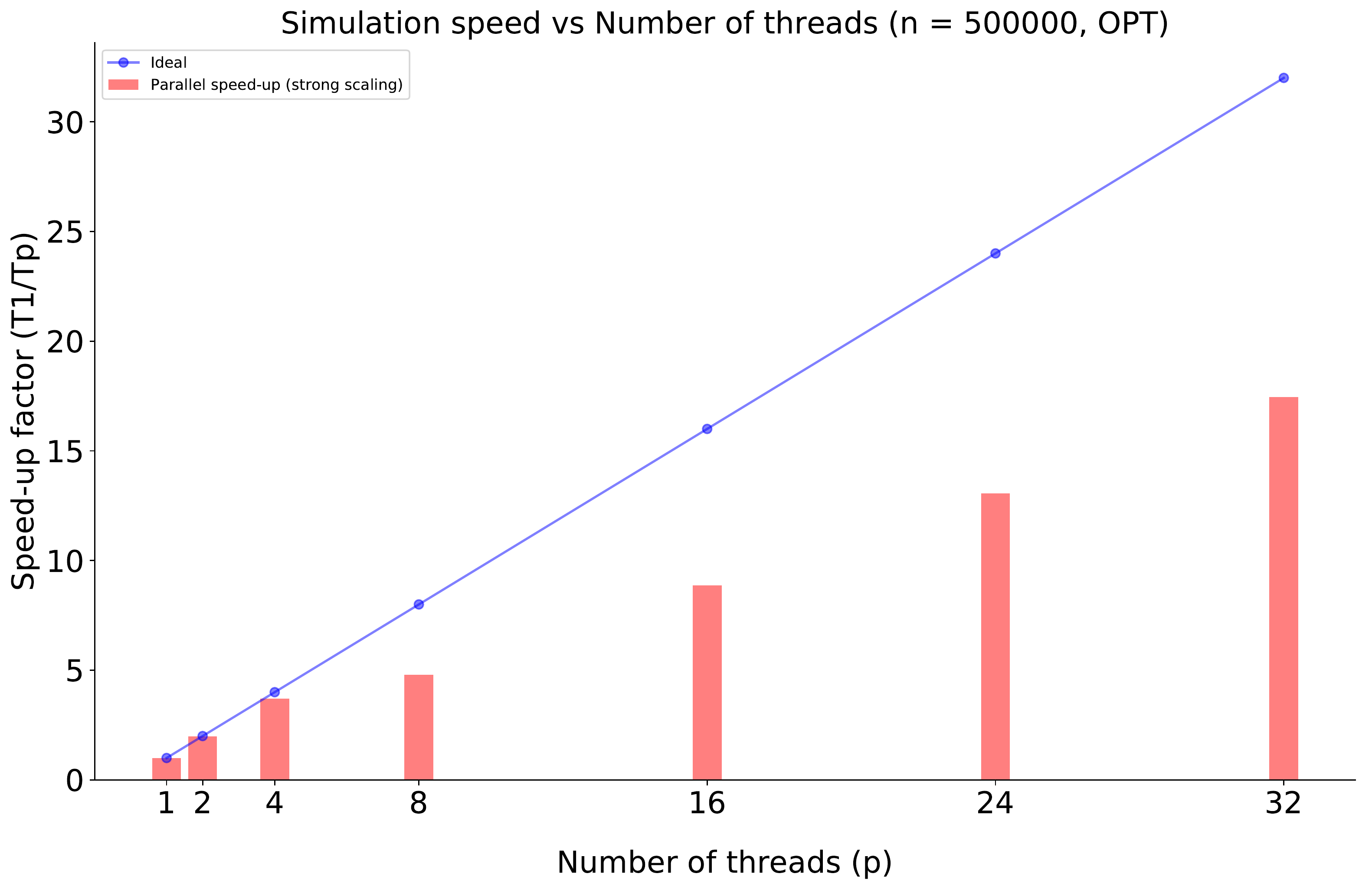}
   \caption{Speedup Factors}
   \label{Speed1}
\end{subfigure}
\caption{Strong-scaling and speedup factors for OPT version $n = 500,000$.}
\label{StrongOPT}
\end{figure}

After generating the speedup and strong-scaling efficiency plots for the experimental instances, we observe that the optimized implementation is able to obtain up to 15x and 20x speedups for the small and large instances respectively, as well as averages strong efficiency factors between 75\% and 82\%, depending on the size and structure of the forest. In Figures \ref{StrongOPT} (a) and (b) we present the results obtained for the average values obtained among 20 instances with 500,000 cells using the optimized OpenMP implementation. Similar --- slightly better --- results are obtained for larger instances.

Based on all our experiments, adding more threads leads to better execution times following a flat pattern w.r.t. the strong-scaling efficiency. Thus, our optimized implementation is able to obtain a great strong scaling performance, taking into account the high complexity of the instances and variability of the results depending on the forest's structure. 

\subsection{Weak Scaling}
Due to the high dependency of the results on forest structure, we divided our analysis for heterogeneous and homogeneous forests. When the forests are heterogeneous, the comparison between instances of different sizes and number of threads looses its meaning since there is no guarantee that the problem will scale in complexity: increasing the problem size doesn't affect the computations time directly, it depends significantly on the composition of the forest, leading to different fire dynamics. Therefore, we expect an erratic pattern when dealing with heterogeneous forests in terms of weak-scaling efficiency.

\begin{figure}[h!]
   \centering
   \includegraphics[scale=0.23]{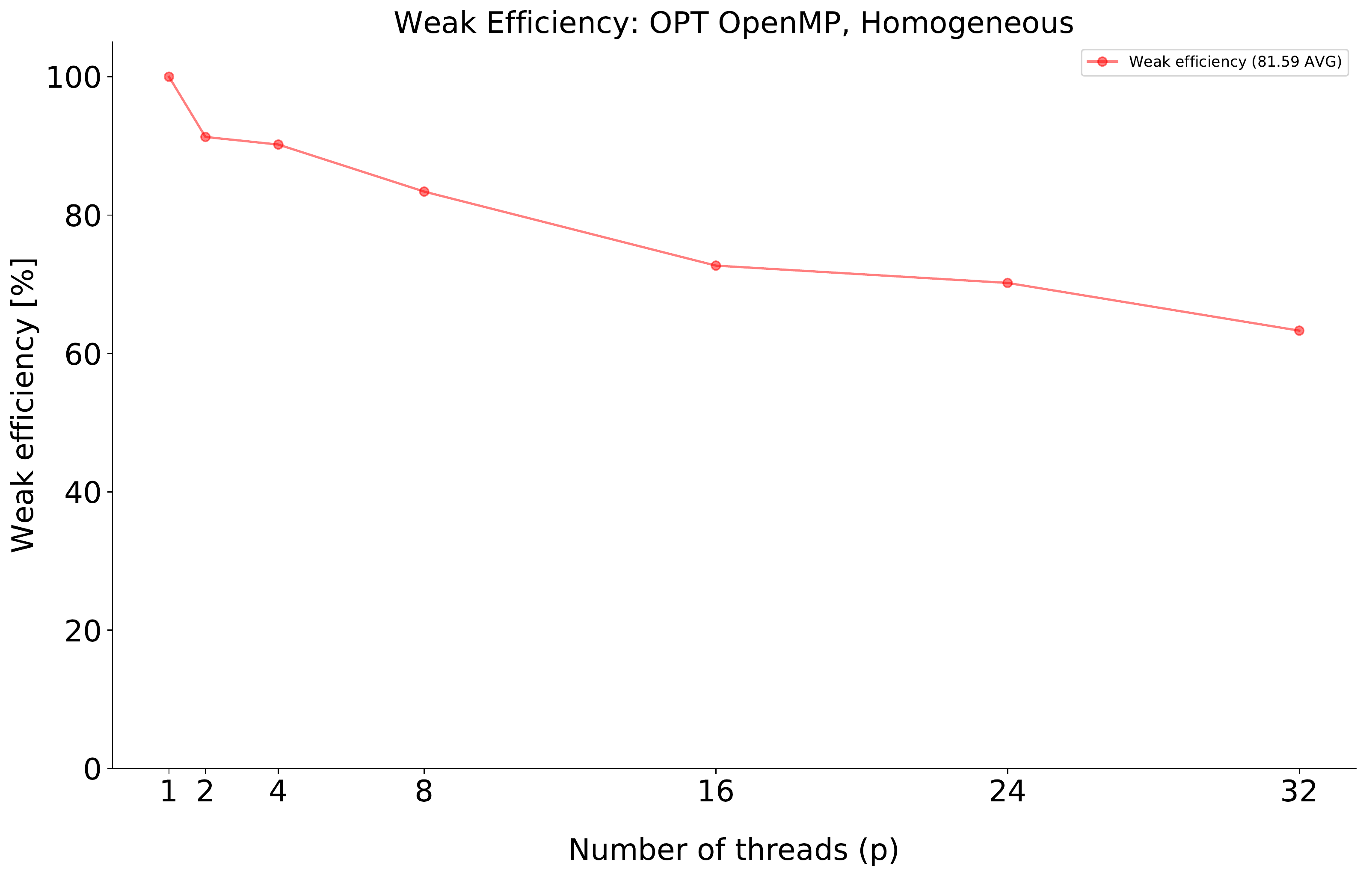}
   \caption{Weak-scaling efficiency (Homogeneous)}
   \label{Weak2}
\caption{Weak-scaling and speedup factors for homogeneous instances, starting with $n = 500,000$.}
\label{Weak}
\end{figure}
In Figure \ref{Weak} we can see the weak efficiency obtained for homogeneous instances. As expected, results for heterogeneous instances lack meaning since we are not comparing the same fire dynamics (and thus, the number of simultaneous burning cells, critical for the parallel performance). On the other hand, comparing homogeneous instances gives us correct results, since we compare the same fire dynamics reaching an average weak-scaling factor value equal to 81.6\%. Again, similar and even better weak factors are obtained with larger (and homogeneous) instances, following the discussion above.

Therefore, our optimized parallel implementation is able to obtain high-performance values in both strong and weak scaling factors thanks to its naturally parallel design.

\end{document}